\documentclass[twocolumn]{aastex701}
\usepackage{hyperref}
\usepackage{array}
\usepackage{booktabs}
\usepackage{float}
\usepackage{verbatim}
\usepackage{natbib}
\usepackage{amsmath}
\usepackage{listings}
\usepackage{multirow}
\usepackage{tabularx}
\usepackage{xcolor}

\graphicspath{{./}{figures/}}

\begin{document}

\title{Trajectory-Agnostic Asteroid Detection in TESS with Deep Learning}

\author[orcid=0000-0003-0501-2636]{Brian P. Powell}
\affiliation{NASA Goddard Space Flight Center, 8800 Greenbelt Road, Greenbelt, MD 20771, USA}
\email[show]{brian.p.powell@nasa.gov}

\author[orcid=0000-0002-7395-4935]{Jorge Martinez-Palomera}
\affiliation{University of Maryland, Baltimore County, 1000 Hilltop Circle, Baltimore, MD 21250, USA}
\affiliation{NASA Goddard Space Flight Center, 8800 Greenbelt Road, Greenbelt, MD 20771, USA}
\email{jorge.i.martinezpalomera@nasa.gov}

\author[orcid=0000-0002-2830-9064]{Amy Tuson}
\affiliation{University of Maryland, Baltimore County, 1000 Hilltop Circle, Baltimore, MD 21250, USA}
\affiliation{NASA Goddard Space Flight Center, 8800 Greenbelt Road, Greenbelt, MD 20771, USA}
\email{}

\author[orcid=0000-0002-3385-8391]{Christina Hedges}
\affiliation{University of Maryland, Baltimore County, 1000 Hilltop Circle, Baltimore, MD 21250, USA}
\affiliation{NASA Goddard Space Flight Center, 8800 Greenbelt Road, Greenbelt, MD 20771, USA}
\email{}

\author[orcid=0000-0003-4206-5649]{Jessie Dotson}
\affiliation{NASA Ames Research Center, Moffett Field, CA 94035, USA}
\email{}

\author[orcid=0000-0001-9125-5591]{Jordan Caraballo-Vega}
\affiliation{NASA Goddard Space Flight Center, 8800 Greenbelt Road, Greenbelt, MD 20771, USA}
\email{}

\correspondingauthor{Brian P. Powell}

\received{24 February 2026}
\revised{13 April 2026}
\accepted{11 May 2026}

\begin{abstract}
We present a novel method for extracting moving objects from TESS data using machine learning. Our approach uses two stacked 3D U-Nets with skip connections, which we call a W-Net, to filter background and identify pixels containing moving objects in TESS image time-series data. By augmenting the training data through rotation of the image cubes, our method is robust to differences in speed and direction of asteroids, requiring no assumptions for either parameter range which are typically required in ``shift-and-stack'' type algorithms. We also developed a novel method for learned data scaling that we call Adaptive Normalization, which allows the neural network to learn the ideal range and scaling distribution required for optimal data processing. We built a code for creating TESS training data with asteroid masks that served as the foundation of our effort (\texttt{tess-asteroid-ml}), which we publicly released for the benefit of the community. Our method is not limited to TESS, but applicable for implementation in other similar time-domain surveys, making it of particular interest for use with data from upcoming missions such as the Nancy Grace Roman Space Telescope and NEOSurveyor.

\end{abstract}


\section{Introduction}\label{sec:intro}

NASA's Transiting Exoplanet Survey Satellite \citep[TESS,][]{Ricker14} is a space-based all-sky photometric survey that has been active since 2018.
Although its primary science goal is to detect exoplanets via the transit method, its unique instrumental characteristics and observation schedule enable numerous other scientific investigations. 
Thanks to its large field of view, high cadence, and fixed-pointing observations, TESS has observed many thousands of solar system asteroids, particularly near the ecliptic plane. 
In the optical band, asteroid detection is limited by the amount of reflected sunlight, which is determined by the asteroid's size and albedo. 
Systematic optical searches for asteroids have been mainly conducted by LINEAR \citep{2000Icar..148...21S}, the Catalina Real-Time Transient Survey \citep{2009ApJ...696..870D}, Pan-STARRS \citep{2013PASP..125..357D}, the Asteroid Terrestrial-impact Last Alert System \citep[ATLAS;][]{2018PASP..130f4505T}, and recently by the NSF-DOE Vera C. Rubin Observatory \citep{2019ApJ...873..111I}. 
In contrast, detection in the infrared is due to the asteroid's thermal emission. 
Most infrared discoveries were made by NASA's WISE and NEOWISE missions \citep{2011ApJ...741...68M, 2011ApJ...743..156M}, which concluded in 2024. 
This effort will continue with NASA's upcoming NEO Surveyor mission \citep{2023PSJ.....4..224M}, expected to launch no later than 2027, which is tasked with finding at least two-thirds of near-Earth objects (NEOs) larger than 140 meters, including potentially hazardous asteroids (PHAs).

\citet{2018PASP..130k4503P} discussed the prospects for solar system science using TESS data, comparing its expected photometric precision to the previous Kepler/K2 mission and analyzing phased light curves for asteroid rotations. Systematic time-series analyses of known asteroids have been performed mostly with TESS data from Years 1 and 2. \citet{2020ApJS..247...26P} analyzed and published light curves from almost 10,000 main-belt and Trojan asteroids observed during Year 1. 
\citet{2023AJ....166..152M} systematically extracted light curves for more than 37,000 asteroids using data from Years 1 and 2. 
Both works performed rotational period estimations for multiple asteroid families, and \citet{2023AJ....166..152M} used multi-sector light curves to study asteroid spin and derive object shapes. 
Past work has focused mostly on the 30-minute cadence data from the TESS prime mission, which enables fainter detections. 
In subsequent extended missions, observations have been made at higher cadences (10-minute and 200-second). 
This decreases the detection limit for single frames but enables analyzing shorter periods of variability. 
Furthermore, during its extended missions, TESS observed sectors that cover gaps left by the prime mission, particularly over the ecliptic plane. 
With the third extended mission, TESS has covered nearly the entire sky, enabling survey data without selection biases. 
A complete asteroid time-series analysis using all available TESS data remains to be done.

Image-based detection algorithms often rely on stacking methods to increase the signal-to-noise ratio (SNR) for faint objects \citep[e.g.][]{2014ApJ...792...60Z}. 
These algorithms—often referred to as shift-and-stack, digital tracking, or track-before-detect—operate on two main assumptions: (1) the asteroid's motion is approximately linear over short timescales relative to the survey cadence, and (2) the asteroid's direction and speed are within a predefined range of parameters. 
In these algorithms, consecutive image frames are shifted by a certain number of pixels given by a proposed speed in a trajectory given by a proposed direction of motion. 
The shifted images are then stacked, increasing the SNR for objects whose projected tracks match the proposed parameters, while background stars are blurred. 
Although these assumptions lead to efficient algorithms and fainter detections, they are inherently limited by the defined parameter range. 
This can lead to missing objects on peculiar trajectories, such as those in high-inclination orbits, those exhibiting apparent changes in direction, or other fast-moving objects.

The detection of faint moving objects in TESS data has been explored through a range of approaches with varying goals, methods, and constraints, making direct comparison between them inherently challenging.  The application of digital tracking to TESS data was first proposed by \citet{2019RNAAS...3..160H}, who suggested a 50\% detection efficiency at magnitude $I_c=22.0$ (despite a wider bandbass, TESS is centered on the $I_c$ band, making the comparison approximately equivalent). This was later revised by \citet{2019RNAAS...3..172P} to $I_c=21.5$ after modeling stray light and stellar contamination. The first study to apply shift-and-stack to TESS was conducted by \citet{2020PSJ.....1...81R}, who applied the method in TESS sectors 18 and 19 in a search for the hypothesized orbit of Planet Nine. 

\citet{2021PASP..133a4503W} developed a single-frame detection pipeline for asteroids in TESS full-frame images.  Rather than employing shift-and-stack techniques, their approach detects point sources in individual difference images then links per-frame detections into tracks.  Comprehensively applying their method to all TESS data from Years 1-2, the authors discovered thousands of new new asteroid tracks reported to the Minor Planet Center (MPC).  Since the \citet{2021PASP..133a4503W} pipeline operates on individual frames, its sensitivity limit (90\% complete at $V\approx19$) demonstrates the single-frame detection threshold of the TESS instrument and is not directly comparable to higher limits achievable through multi-frame detection, which improves the signal-to-noise ratio in proportion to the square root of the combined number of frames.

More recently, \citet{2024AJ....167..113N} introduced a more computationally-efficient implementation of the shift-and-stack method using the fast discrete X-ray transform (FaXT). 
This algorithmic improvement reduced the number of operations from $\mathcal{O}(N^5)$ to 
$\mathcal{O}(N^4)$, where $N$ represents the approximate size of each dimension (image size, number of frames, and tested velocities).  The authors applied the FaXT pipeline to a subset of TESS Sector 5 data with a 128-frame stack and achieved 40\% completeness at $V\approx21$, broadly consistent with the predictions of \citet{2019RNAAS...3..172P}, while differences in stack depth, search volume, and noise conditions make precise comparisons difficult. \citet{2025AJ....170..187Z} proposed an improvement to multi-frame shift-and-stack whereby a gradient-based algorithm can increase the SNR of a stacked image by iteratively changing the velocity hypothesis in the direction of a higher SNR, reducing the need for comprehensive velocity hypotheses in the initial search.

In this work, we develop an image-based detection approach free from the constraints of speed or trajectory hypotheses. To accomplish this, we use a machine learning (ML) method to identify moving objects in TESS data. ML has previously been used for asteroid detection purposes, although not at the pixel-level. For example, \citet{2021PASP..133c4501C} developed a convolutional neural network to classify whether four consecutive Asteroid Terrestrial-impact Last Alert System (ATLAS) images contained an asteroid track. 
\citet{COWAN2023100693} used a similar technique to classify image stacks from the Microlensing Observations in Astrophysics (MOA) survey.

Our model uses a 3D Convolutional Neural Network (CNN), an architecture naturally suited for this task as it can learn spatio-temporal features simultaneously, identifying motion patterns in a sequence of images without prior assumptions about direction or speed. 
The model outputs a probability map indicating where an asteroid is located in both image and time dimensions. TESS Full Frame Images (FFIs) provide the perfect proving grounds for this method, though this approach could be adapted to other time-domain surveys, such as NASA's upcoming Nancy Grace Roman Space Telescope (Roman) and NEO Surveyor missions.

The arrangement of this paper is as follows: in Section \ref{sec:data_prep}, we discuss the TESS and JPL Horizons data and detail the development of our publicly-available code for data preparation. 
We then discuss the structure of the neural network in Section \ref{sec:nn}, with our Adaptive Normalization method detailed in Section \ref{sec:can}. Section \ref{sec:loss} describes our model loss function, while Section \ref{sec:training} outlines the model training procedure. 
Our results are shown in Section \ref{sec:results}, followed by a discussion in Section \ref{sec:discussion}. 
Finally, we summarize and provide our conclusions in Section \ref{sec:conclusions}.

\section{Data Preparation}
\label{sec:data_prep}

In this section, we detail our efforts to build the training data. This led to the development of \texttt{tess-asteroid-ml} \citep{tess-asteroid-ml} -- a publicly available Python package to assist the community with further investigations of ML methods for detecting asteroids in TESS data.

\subsection{TESS Data}
\label{subsec:tess_data}

TESS observes the sky with four cameras, creating a combined field of view (FOV) of $96 \times 24$ deg.
Each camera has a $24 \times 24$ deg FOV and contains 4 CCDs arranged in a $2 \times 2$ layout.
The science portion of each CCD has an array of $2048 \times 2048$ pixels with a pixel scale of $21 \arcsec / \text{pix}$.

\begin{figure}[htb!]
    \centering
    \epsscale{1.15}
    \plotone{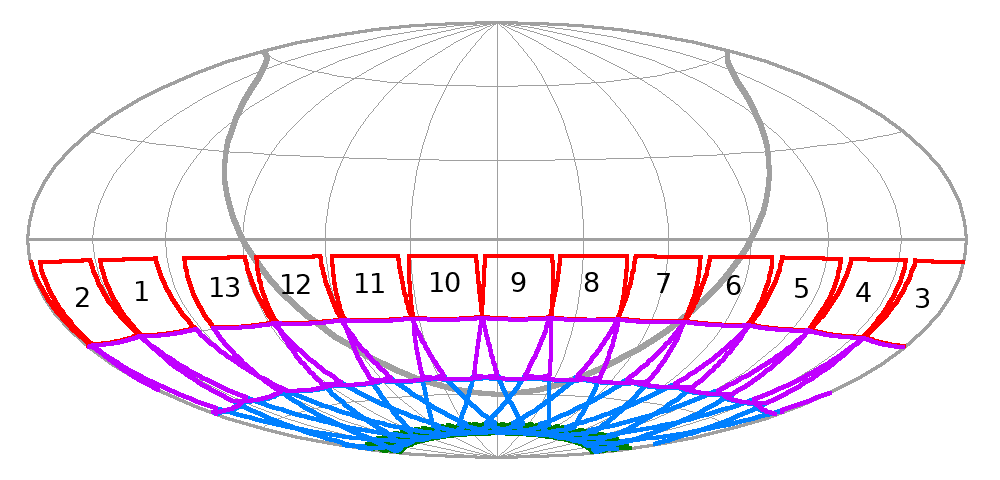}
    \plotone{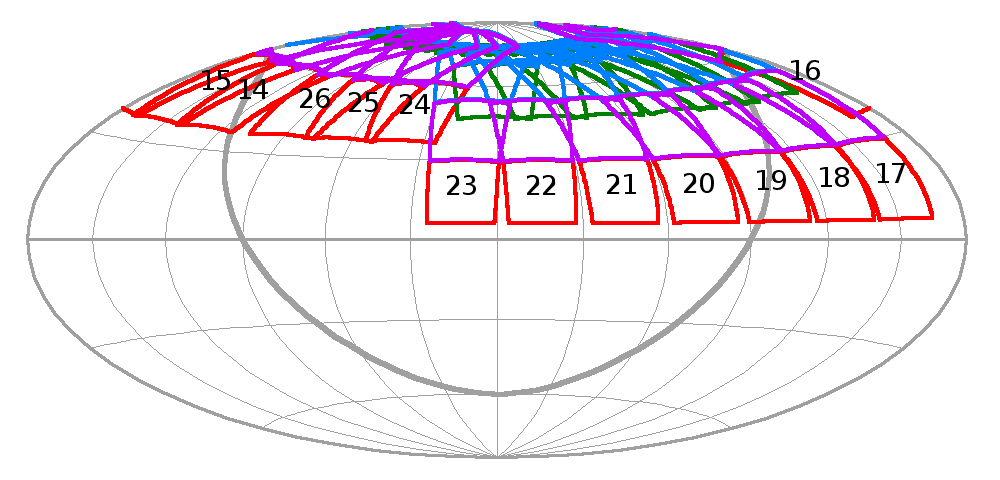}
    \caption{
    TESS Year 1 (top) and 2 (bottom) sky coverage from \url{https://tess.mit.edu/observations/}. 
    The figure is projected in ecliptic coordinates with the Galactic plane highlighted in bold gray. Sector numbers are noted in each pointing. The instrument cameras are highlighted with different colors, camera 1 in red, camera 2 in purple, camera 3 in blue, and camera 4 in green. Sectors 14-16 and 24-26 are shifted away from the ecliptic plane to reduce strong scattered light. 
    }
    \label{fig:tess_sectors}
\end{figure}

During its primary mission, TESS observed the southern hemisphere (Year 1, Sectors 1-13) and the northern hemisphere (Year 2, Sectors 14-26). 
In each sector, the spacecraft maintained a stable pointing for approximately 27\,days. 
The observing strategy was designed to place Camera 1 near the ecliptic plane, with the others vertically stacked towards the ecliptic pole. 
Some sectors in the northern hemisphere were offset from the ecliptic to reduce stray light. Figure \ref{fig:tess_sectors} shows the TESS sector map in ecliptic coordinates.
FFIs from the primary mission were collected at a 30-minute cadence, producing roughly 1300 frames per sector. 

We used the calibrated TESS FFIs \citep{https://doi.org/10.17909/0cp4-2j79}, which are publicly available through the Mikulski Archive for Space Telescopes (MAST). 
They can be accessed as single-frame FITS files or as data cubes $(N_t\times2048\times2048)$ on Amazon Web Services (AWS). 
We accessed the data using \texttt{tess-cube}\citep{Hedges_tesscube}, which enables efficient retrieval of specific sections of the FFI cubes from the cloud without requiring local copies of the entire dataset. 
The latter is important, as a single FFI cube for a given sector, camera, and CCD is about 50 GB, making bulk storage difficult.

\subsection{Asteroid Catalog and Ephemeris}
\label{subsec:asteroid_data}

To build a training set, we needed ground-truth labels of known asteroid trajectories visible in the TESS data. A natural starting point for creating these labels was
the NASA Jet Propulsion Laboratory (JPL) Horizons system, which provides ephemerides for over 1.3 million asteroids and other solar system bodies.

\begin{figure}[htb!]
    \centering
    \epsscale{1.18}
    \plotone{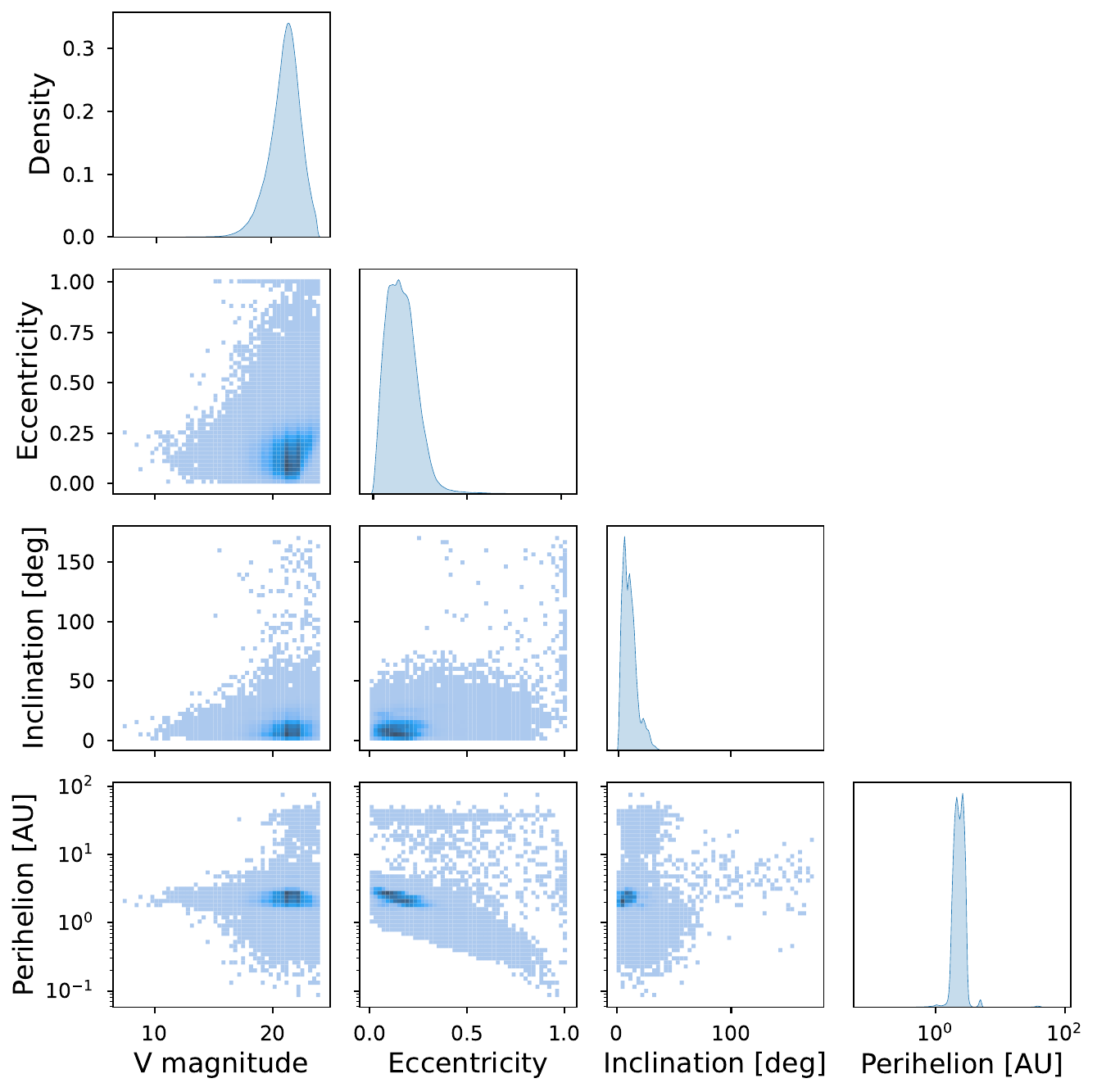}
    \caption{
    Joint distribution of known asteroids observed during TESS Years 1 and 2. The magnitude range reflects the limiting magnitude of $V < 24$ used when querying JPL Horizons. The orbit inclination values are dominated by asteroids near the ecliptic plane. The peak in the distribution of perihelion values correspond to main-belt asteroids.
    }
    \label{fig:asteroid_input_catalog}
\end{figure}

Using the Small-Body Database API, we first obtained a preliminary list of asteroids and comets down to magnitude $V=25$ potentially visible in each TESS sector. 
Then, for each object, we used the Horizons API to generate precise ephemeris data using TESS spacecraft time and state vectors (position and velocity) as the observer in order to account for parallax effects.
We converted the resulting Right Ascension (R.A.) and Declination (Decl.) coordinates to pixel row and column positions using the World Coordinate Solution (WCS) from the FFI headers. 
We only retained asteroid trajectories with an expected mean apparent magnitude brighter than $V<24$. 
This limiting magnitude was selected to ensure our label set is complete well beyond the known TESS detection limit of $V\approx21$ \citep{2019AJ....158..138S}, which prevents biasing our training data and provides flexibility for experimenting with different magnitude limits. 
Figure \ref{fig:asteroid_input_catalog} shows the joint distribution of orbital elements for the resulting asteroids.

\begin{figure}[ht!]
    \centering
    \epsscale{0.7}
    \plotone{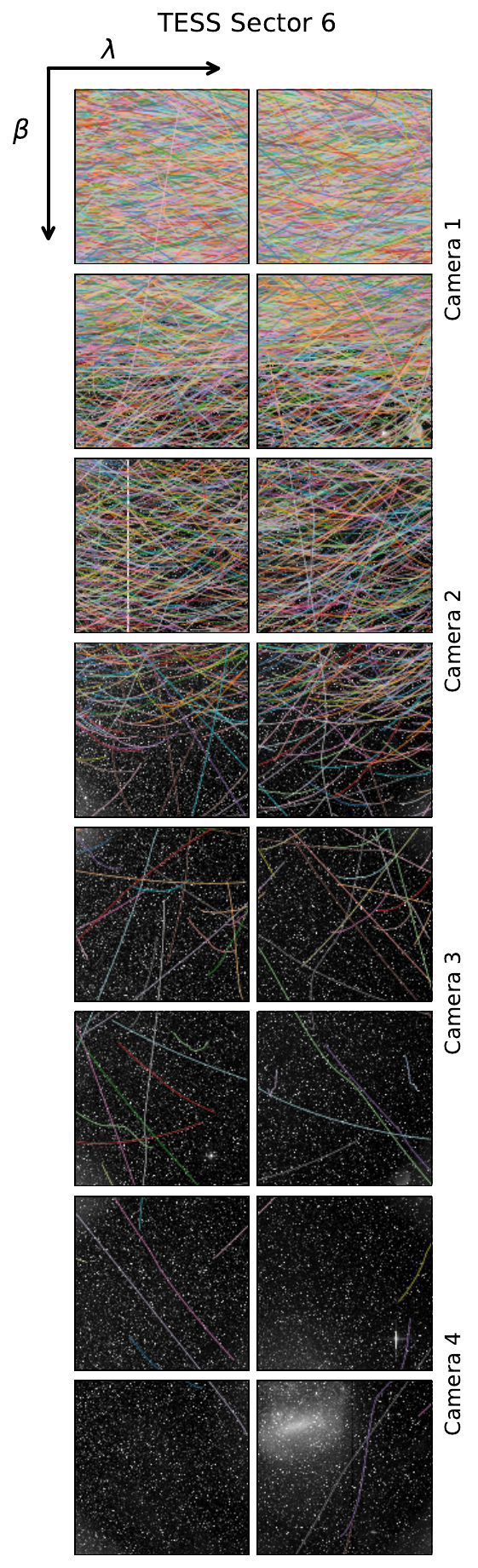}
    \caption{Tracks of asteroids observed during TESS Sector 6 (2018/12/11 - 2019/01/07). Each panel shows a TESS FFI from one of the 16 CCDs. The colored lines represent the asteroid tracks, where the colors have no meaning other than representing different asteroids. The higher density of tracks in camera 1 (top two rows) is due to its proximity to the ecliptic plane, where most of the Solar System's minor bodies orbit. Asteroids with highly inclined orbits are visible towards polar latitudes.
    }
    \label{fig:jpm_sector_tracks}
\end{figure}

From our queries, a typical TESS sector contains $\sim\!30,000$ unique asteroid tracks. 
Figure \ref{fig:jpm_sector_tracks} shows the projected tracks in Sector 6. 
Due to the TESS instrument layout, Camera 1, which is closest to the ecliptic plane, contains over 90\% of the observable asteroids in a sector. 
Camera 2 typically contains $\sim8\%$, while Cameras 3 and 4 account for the remaining $\sim2\%$ of asteroids. 
This follows the characteristic distribution of minor bodies away from the ecliptic plane. 
The majority of these asteroids have low-inclination orbits, appearing as horizontal tracks. 
Minor bodies with higher inclinations project as diagonal or vertical tracks and are more common in Cameras 3 and 4.

Across the southern hemisphere (Sectors 1-13), we cataloged $259,602$ asteroid tracks. 
For the northern hemisphere (Sectors 14-26), we found $150,865$ tracks. 
The lower count in the north is due to the six sectors that were pointed away from the ecliptic plane (see Figure \ref{fig:tess_sectors}). 
Combining both hemispheres, we account for $401,467$ tracks from $364,036$ unique asteroids.

\subsection{Data Preprocessing}
\label{subsec:data_prep}

We developed the \texttt{tess-asteroid-ml} Python package to process the TESS FFIs and asteroid tracks into a format suitable for training our neural network model. 
The pipeline produces four data products for each processed region of the sky:
\begin{enumerate}
    \item \textbf{Image cutout}: A $64\times64$ pixel cutout of the FFI cube shaped into a 3D data array with the flux image for every observed cadence. We subtracted each frame with a median-frame computed from the same 64-size cubes. The image cutouts contain flux values in units of $e^{-}/s$ and the shape is $[N_{\text{cadences}}, 64, 64]$.
    \item \textbf{Asteroid mask}: A corresponding 3D boolean mask of shape $[N_{\text{cadences}}, 64, 64]$ marking the pixels where an asteroid is present according to its ephemeris. To create a mask for each asteroid, we generated elliptical apertures around the projected ephemeris for each frame. The ellipse shape is proportional to the object's projected speed, and its size is scaled by a power-law function of the object's expected magnitude. The mask has a value of 1 inside the aperture and 0 elsewhere. We then combined each individual mask to create the global asteroid mask.
    \item \textbf{Asteroid Catalog}: A table listing the asteroids within the cutout and their metadata (e.g., magnitude, orbital elements).
    \item \textbf{Metadata}: A file containing time values, cadence numbers, and the reference pixel coordinates of the cutout on the full CCD.
\end{enumerate}

We acknowledge two minor simplifications in our coordinate conversions when creating the asteroid masks: we did not account for spacecraft velocity aberration or the full barycentric time correction for pixels far from the camera center. 
These effects introduce small positional offsets, resulting in label noise (i.e., mask pixels not perfectly aligned with flux). 
However, during development, we found that the network successfully generalizes over this noise (see Section \ref{sec:completeness}), so we omitted these corrections to simplify the data pipeline.

To process an entire FFI cube, we generated a grid of $64\times64$ pixel cutouts, with a 4-pixel overlap to mitigate edge effects, and 64 frames in the time axis. 
At a 30-minute cadence, 64 frames cover 32 hours, long enough to establish motion but short enough to keep memory requirements manageable.
This temporal slicing accounts for data gaps (e.g., downlinks) and allows the model to make multiple predictions for each part of an asteroid's track, enhancing robustness.
This results in 1,156 cutout/mask pairs per CCD. 
These 64-size minicubes with the image cutouts and asteroid masks are the data used for model training.

We built these 64-size data cubes for all of TESS Years 1-2 (Sectors 1-26), including all known asteroid tracks from JPL Horizons brighter than $V<22$\,mag. We discuss the selection of this magnitude cut further in Section \ref{sec:magnitude_lim}. Construction time for this data is variable and depends linearly on the number of asteroid tracks. On average, all the data cubes for a single sector/camera/CCD combination will take 2-3 hours to build on a single CPU, with particularly dense or sparse CCDs taking slightly longer or shorter, respectively. 
The Python package used to build this training set is publicly available as \texttt{tess-asteroid-ml}\footnote{Code, documentation, and tutorials available at \url{https://github.com/jorgemarpa/tess-asteroid-ml}} to ensure the reproducibility of our results.
The \texttt{tess-asteroid-ml} code is discretized at the CCD level, so this process is easily parallelizable.

\section{Model Development}
\label{sec:nn}

Our model needed to transform input data (the flux cubes described in Section \ref{sec:data_prep}) into a mask that indicated the per-pixel probability of the presence of an asteroid. Broadly speaking, the problem of providing input data to a neural network and creating an output mask is known as semantic segmentation. In semantic segmentation, the neural network learns to identify characteristics of an object in an image. In the simplest form with a single class of object, the mask created by the neural network is then a map of those pixels containing the target object (with a score of 1) and pixels not containing the target object (with a score of 0). The most common type of neural network used to accomplish the task of semantic segmentation is called a U-Net \citep{10.1007/978-3-319-24574-4_28}.

The name of the U-Net is derived from its appearance as a ``U''. That is, an input image is decomposed through multiple layers and then reconstructed. Generally, such a progression is associated with an autoencoder type neural network, originally proposed by \citet{lecun-87}. However, the key characteristic making a U-Net distinct from an autoencoder is the existence of ``skip connections.'' These connections pass information directly across from one branch of the ``U'' to the other, providing alternative paths for the flow of the gradient during training and for the preservation of structural information about the image, which would otherwise be lost in the dimensional reduction of the innermost layers. See Figure 1 of \citet{10.1007/978-3-319-24574-4_28} for a diagram of a typical U-Net.

U-Nets have been successfully employed in a variety of applications \citep[e.g.][]{LAN2020100197,diagnostics11030501,s21093179,icaart23,10.1145/3654522.3654573}.  As in their original development for biomedical image segmentation by \citet{10.1007/978-3-319-24574-4_28}, much of the development and innovation has taken place in this field, such as the combination of U-Nets with vision transformers \citep{CHEN2024103280}.  U-Nets have also driven substantial progress in stable diffusion models \citep{9878449}, which originally used U-Nets as an internal denoising mechanism.

\subsection{Neural Network Architecture}

We started our examination of the TESS data cubes using the concept of the U-Net. Our task, however, had the additional difficulty of the added temporal dimension. To accommodate the temporal dimension, the U-Net needed a structure with 3D convolutions rather than traditional U-Net structures with 2D convolutions. The convolutional filters would then need to learn to identify temporal differences in the position of a light source. The filters would also need to develop a means to express that stationary light sources are unimportant to the classification.  

\begin{figure*}
    \centering
    \includegraphics[width=.95\linewidth]{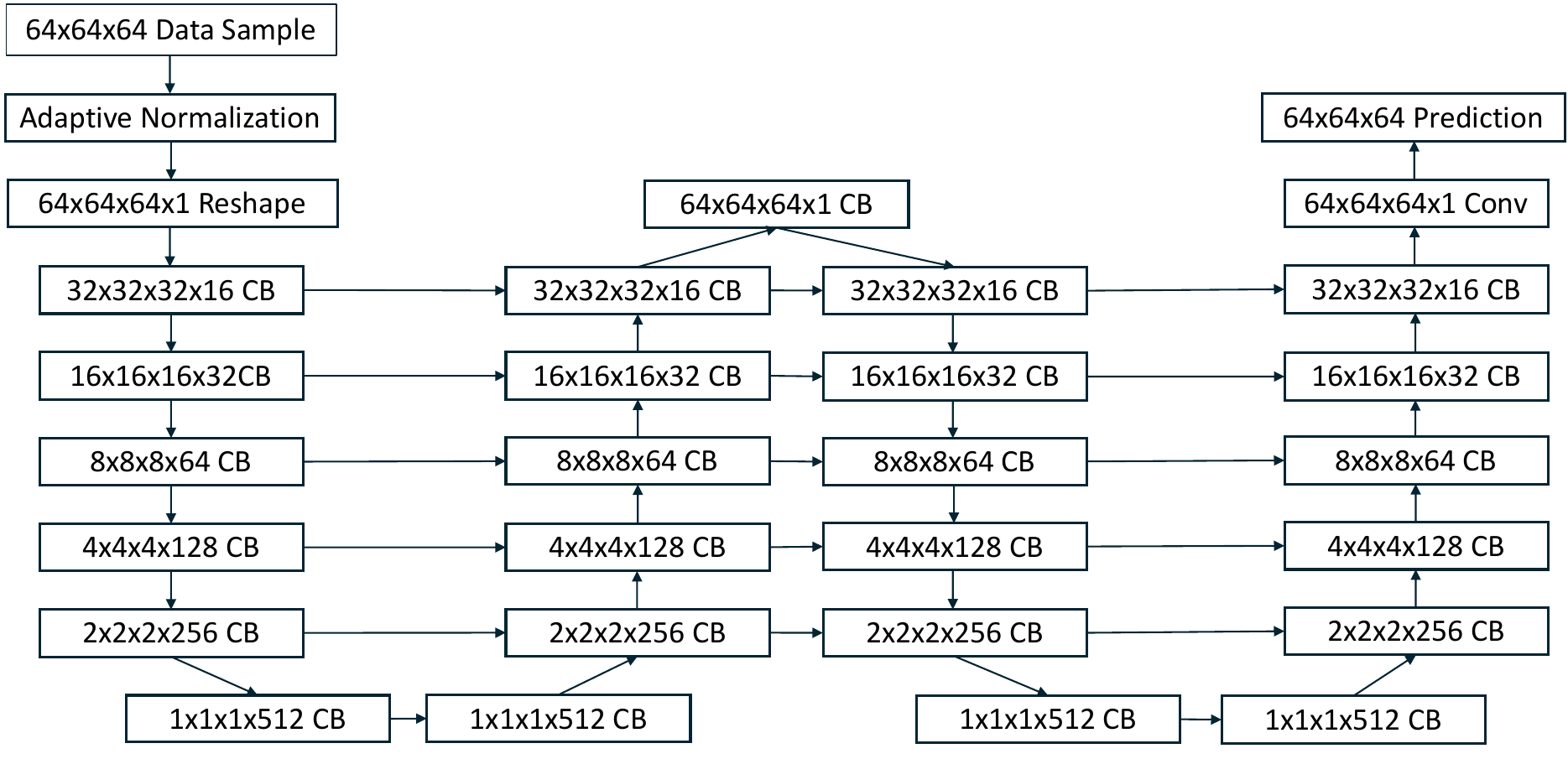}
    \caption{W-Net structure. A 64$\times$64$\times$64 sample of median-subtracted TESS data built using the process described in Section \ref{sec:data_prep} is the input to the neural network. The data undergoes our custom Adaptive Normalization process, described in Section \ref{sec:can}, where the model learns the parameters of a mixture of Logistic cumulative distribution functions (CDFs) that provide a learned transformation of the input data to the [0,1] range. Convolutional blocks (CBs) are given by their output dimensions and described further in Figure \ref{fig:cb}. Activations internal to the convolutional blocks use the Exponential Linear Unit (ELU) function \citep{clevert2016elu}, while the final and innermost 64$\times$64$\times$64$\times$1 convolutions are activated with the sigmoid function.}
    \label{fig:wnet}
\end{figure*}

\begin{figure}
    \centering
    \includegraphics[width=.75\columnwidth]{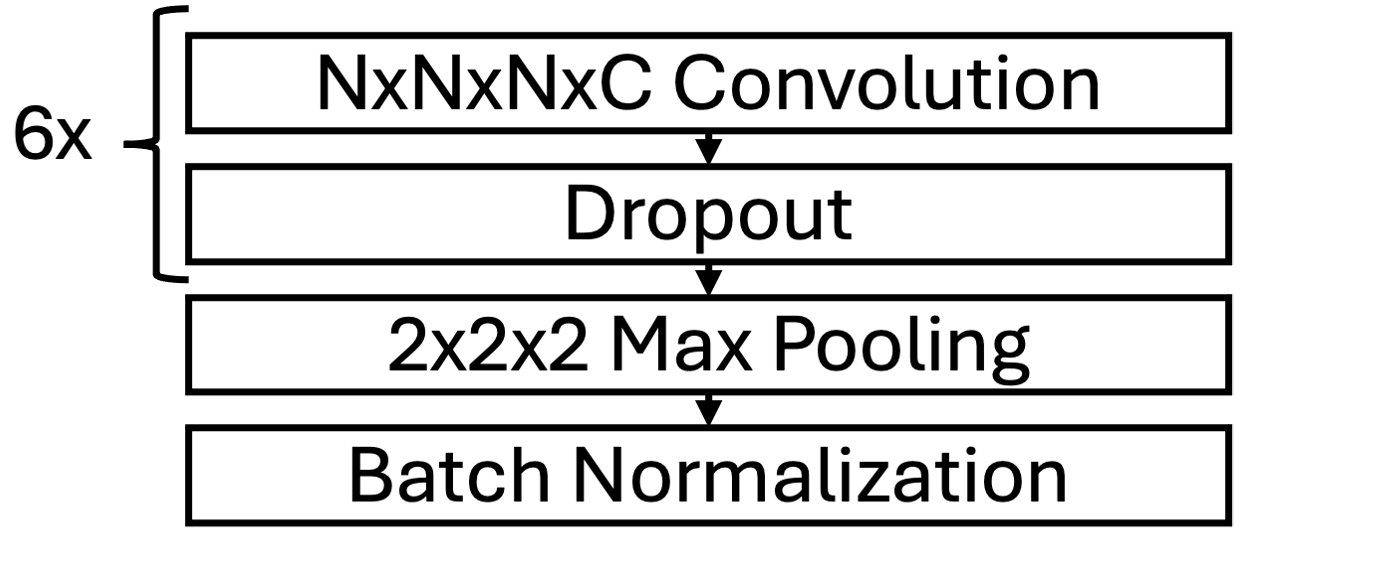}
    \caption{Structure of a convolutional block from the W-Net. Each convolutional block of dimension N with C channels contains six 3$\times$3 convolutional layers alternated with dropout, followed by a max pooling layer with a 2$\times$2$\times$2 kernel to reduce the dimensionality, finishing with a batch normalization layer. Dropout fractions are 0.1 for the 16, 32, and 64 channel convolutional layers, 0.2 for the 128 and 256 channel convolutional layers, and 0.3 for the 512 channel convolutional layers. Concatenations for skip connections are made in the channel dimension.}
    \label{fig:cb}
\end{figure}

As we developed the network structure using the {\sc tensorflow} \citep{tensorflow} and {\sc keras} \citep{keras} python libraries and began to test performance, we noticed that our 3D U-Net was only poorly able to identify asteroids. In order to remedy the poor performance, we progressively made the network deeper. Ultimately, rather than continuing to make the U-Net itself deeper, we stacked two U-Nets and created additional skip connections between the upward arm of the first ``U'' and the downward arm of the second ``U,'' effectively making a ``W-Net.'' The W-Net proved to be effective and so we moved forward with this structure. See Figures \ref{fig:wnet} and \ref{fig:cb} for a more thorough description of the structure. We did not conduct a sensitivity analysis of different structures as we considered this to be beyond the scope of our initial study.

\subsection{Model Scope}
\label{sec:scope}

To evaluate our model against the entirety of TESS data from Years 1-2, we needed to train a minimum of two distinct models. Data that was used for training cannot be considered in evaluating the performance of the model, that is, we needed to ensure that the model had no specific knowledge of the data which was used for inference, either in training or validation.\footnote{To avoid confusion, we define three terms as we use them; training, validation, and inference. (i) Training: The process of teaching the model to learn the prediction mask corresponding to respective input samples by providing the truth (the ephemeris mask) -- data used for training cannot be used for validation or inference, since the model will inherently know the output corresponding to the input. (ii) Validation:  The process of evaluating performance of the model (via the loss function described in Section \ref{sec:loss}) -- data used for validation cannot have been used for training. (iii) Inference: The process of using the trained model to predict outputs -- inference cannot be performed on data with which it has been trained or validated.} Another consideration was that maximizing the variety of data used for training improves model generalizability. 

We decided to create two models: one which would be trained and validated on the Southern sectors of TESS Year 1 (Sectors 1-13) and one which would be trained and validated on the Northern sectors of TESS Year 2 (Sectors 14-26). The former (hereafter called the ``ST'' model, for ``South-Trained'') would be used for inference on the Northern sectors, while the latter (hereafter called the ``NT'' model, for ``North-Trained'') would be used for inference on the Southern sectors. In this manner, we ensure that the models were not given any information which could positively skew their performance. Specifically, the ST model was trained on data from TESS Sectors 1, 2, 3, 5, 6, 7, 9, 10, 11, and 13, and validated against Sectors 4, 8, and 12. The NT model was trained on data from TESS Sectors 14, 15, 16, 18, 19, 20, 22, 23, 24 and 26, and validated against Sectors 17, 21, and 25. These choices for training and validation sets were arbitrary and we expect, but did not evaluate, that different selections would have no substantial effect on model performance.  The quantity of available data samples for training and validation are given in Table \ref{tab:model_data}.

Given the wide variety of noise and systematics in TESS data, which is often highly specific, not using data from the same sector/camera/CCD for training/testing and inference presented a substantial challenge for the generalizability of the model. We could have used, for example, different row/column sections of the same sector/camera/CCD combination for training, validation, and inference. In this manner, the model would be trained in the noise or systematics specific to a sector/camera/CCD combination and, as such, performance of the model would likely be improved. However, in order to demonstrate the potential of our method in widely varied conditions, we opted to ensure that the model had no knowledge in training of the entire sector in which it would be used for inference.

\begin{table}
\centering
\begin{tabular}{|c|c|c|r|}
\hline
Model & Sector & Type$^a$ & Samples \\
\hline
\multirow{13}{*}{ST} & 1 & T & 21\thinspace070\thinspace427\\
\cline{2-4}
& 2 & T & 19\thinspace790\thinspace720\\
\cline{2-4}
& 3 & T & 17\thinspace811\thinspace648\\
\cline{2-4}
& 4 & V & 15\thinspace474\thinspace216\\
\cline{2-4}
& 5 & T & 19\thinspace386\thinspace120\\
\cline{2-4}
& 6 & T & 15\thinspace800\thinspace208\\
\cline{2-4}
& 7 & T & 17\thinspace756\thinspace160\\
\cline{2-4}
& 8 & V & 14\thinspace165\thinspace624\\
\cline{2-4}
& 9 & T & 18\thinspace581\thinspace544\\
\cline{2-4}
& 10 & T & 18\thinspace344\thinspace564\\
\cline{2-4}
& 11 & T & 18\thinspace610\thinspace444\\
\cline{2-4}
& 12 & V & 20\thinspace051\thinspace976\\
\cline{2-4}
& 13 & T & 21\thinspace235\thinspace720 \\
\hline
\multirow{13}{*}{NT} & 14 & T & 20\thinspace549\thinspace056 \\
\cline{2-4}
& 15 & T & 19\thinspace679\thinspace744\\
\cline{2-4}
& 16 & T & 18\thinspace348\thinspace032\\
\cline{2-4}
& 17 & V & 18\thinspace105\thinspace272\\
\cline{2-4}
& 18 & T & 17\thinspace604\thinspace724\\
\cline{2-4}
& 19 & T & 18\thinspace939\thinspace904\\
\cline{2-4}
& 20 & T & 19\thinspace489\thinspace004\\
\cline{2-4}
& 21 & V & 20\thinspace876\thinspace204\\
\cline{2-4}
& 22 & T & 20\thinspace515\thinspace532\\
\cline{2-4}
& 23 & T & 19\thinspace810\thinspace372\\
\cline{2-4}
& 24 & T & 20\thinspace308\thinspace608\\
\cline{2-4}
& 25 & V & 19\thinspace254\thinspace336\\
\cline{2-4}
& 26 & T & 18\thinspace865\thinspace920\\
\hline\hline
\multirow{2}{*}{ST} & \multicolumn{2}{c|}{Training total} & 188\thinspace387\thinspace555 \\
\cline{2-4}
& \multicolumn{2}{c|}{Validation total} & 49\thinspace691\thinspace816 \\
\hline
\multirow{2}{*}{NT} & \multicolumn{2}{c|}{Training total} & 194\thinspace110\thinspace896 \\
\cline{2-4}
& \multicolumn{2}{c|}{Validation total} & 58\thinspace235\thinspace812\\
\hline
\end{tabular}
\caption{Number of 64$\times$64$\times$64 samples per sector, with training and validation totals. Notes: (a) T = Training, V = Validation.}
\label{tab:model_data}
\end{table}


\subsection{Data Scaling}
\label{sec:scaling}

Our early attempts at model development included using normalization of the data cubes to [0,1] as input to the neural network, which failed to converge. TESS FFI flux data can have an enormous range, varying from negligible to fully saturated pixels, numerically from $\sim$0 to $\sim$10$^{5}$. Although median subtraction reduces that range, the neural network was unable to effectively extract the relevant values. The need for a better type of scaling was evident.

Data scaling can be essential to ML applications, particularly neural networks. Keeping data within a limited range allows a neural network to focus on finding patterns rather than on mitigating the effects of irrelevant numerical interrelationships. Although there are many different methods for scaling data, we chose to proceed with quantile transformation, which uses a nonlinear transformation to map the probability density distribution to a Uniform Distribution.  This scaling method has proven effective in previous ML applications with TESS data, such as finding eclipses in TESS light curves (see e.g., \citealt{2021AJ....161..162P,2021AJ....162..299P,2022ApJ...938..133P,2025ApJ...985..213P,2022ApJS..259...66K,2024MNRAS.527.3995K,2025ApJS..279...50K,2026AJ....171...29K}).

Quantile transformation of the input data cubes allowed our model to successfully converge. However,  this step was a substantial bottleneck in our process. Quantile transformation was effective but computationally slow and, perhaps more importantly, is something of a blunt instrument as there is no guarantee of an optimal data distribution. We wanted to scale the data effectively such that all the convolutional operations could focus on extracting the asteroids rather than data normalization.

\section{Adaptive Normalization}
\label{sec:can}

In order to address this problem, we sought to create a {\it learned scaling} mechanism internal to the neural network that could adapt to the data distribution. This operation would be conducted in the initial layer of the neural network, performing the role of data scaling that would otherwise be performed as a pre-processing step, but in a learned capacity. 

Normalization layers are not a new concept for neural networks. Batch normalization \citep{ioffe2015batch}, which is commonly used as means of scaling activations of convolutional layers to control the gradient flow, is ubiquitous in CNNs. Layer Normalization \citep{ba2016layer} normalizes across the feature dimension rather than the batch dimension, finding utility in transformer architectures \citep{vaswani2017attention}.  Our method is substantially different, however, in that we fundamentally transform the data itself as the {\em initial} layer.  In this manner, we perform a step usually done in data pre-processing as part of the neural network and, rather than a one-size-fits-all transformation of common scaling methods, we {\em learn} a nonlinear transformation to intelligently optimize the data distribution for follow-on processing by the W-Net.  In this role, our normalization layer is perhaps more similar to the concept of attention \citep{bahdanau2015attention, luong2015effective}, in which features of the data are emphasized or de-emphasized through a learned process. However, our concept is much different from attention in that it emphasizes regions of a distribution rather than individual elements of the data.

Our initial scaling layer needed to (i) input the raw data, (ii) learn a transformation, and (iii) output the transformed data in the range [0,1]. The requirement of an output range of [0,1] is not a necessity for the neural network, but rather our own imposition after seeing the initial success of the neural network with quantile-scaled data in the same range. However, this also guided our selection of the process of transforming the data as a weighted mixture of cumulative distribution functions (CDFs). We specifically chose the CDF of the Logistic Distribution. We do not assert that this distribution would perform any better or worse than another CDF with the same input range, as we chose it only for its simplicity. The role of the CDF in our case is simply as the basis for a transformation function through a weighted combination. For our learned transformation, then, for each CDF, the neural network needed to learn three parameters:  location ($\mu$), scale ($s$), and weight ($w$).

We initialized these parameters for each respective CDF as a learnable weight. That is, each as a variable of length $N_{\text{distributions}}$, the user-defined number of Logistic CDFs in the mixture. Each of the parameters was learnable on an unlimited range. Distribution weights were activated using the softmax function, i.e.

\begin{equation}
w(x)_i = \frac{e^{x_i}}{\sum_{j=1}^K e^{x_j}},
\end{equation}
where $x_{i}$ is an individual learned parameter and $w(x)_i$ is the activated distribution weight, such that the sum of the weights is equal to unity. Also, since the $s$ value of the Logistic Distribution is strictly positive, they were activated using the softplus function, i.e.

\begin{equation}
s(x) = \ln(1 + e^x),
\end{equation}
where $x$ is the learned parameter and $s$ is the activated scale, such that all scale values are positive. Locations, $\mu$, were not activated in order for $\mu$ to have the ability to be positive or negative and adapt to the data distribution as needed (for clarity, we remind the reader that the input to the neural network is the median-subtracted flux image, so can be negative). A loss term was added to encourage $\mu$ to be different for each distribution, improving the richness of the representation. We called this loss term the ``variance loss,'' $L_{v}$, given by

\begin{equation}
L_{v}(\mu) = w_{\mu}\frac{1}{1+e^{-\sigma^{2}(\mu)}},
\label{eqn:lv}
\end{equation}
which is the sigmoid activated variance of the vector $\mu$ multiplied by an arbitrary weight, $w_{\mu}$. The sigmoid activation function is on the scale (0,1), so the user-defined weight determines the importance of the term to the overall model loss function, which we will discuss further in Section \ref{sec:loss}.

\begin{figure}
    \centering
    \includegraphics[width=.99\linewidth]{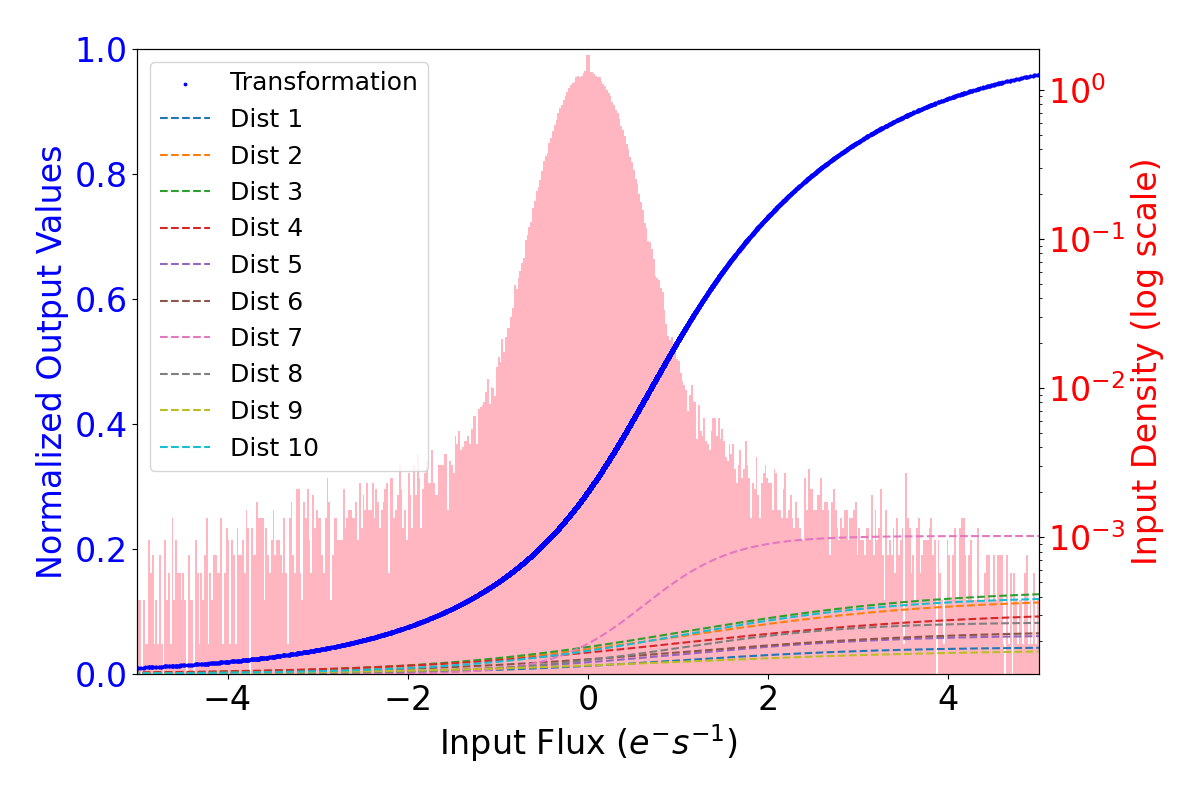}
    \includegraphics[width=.99\linewidth]{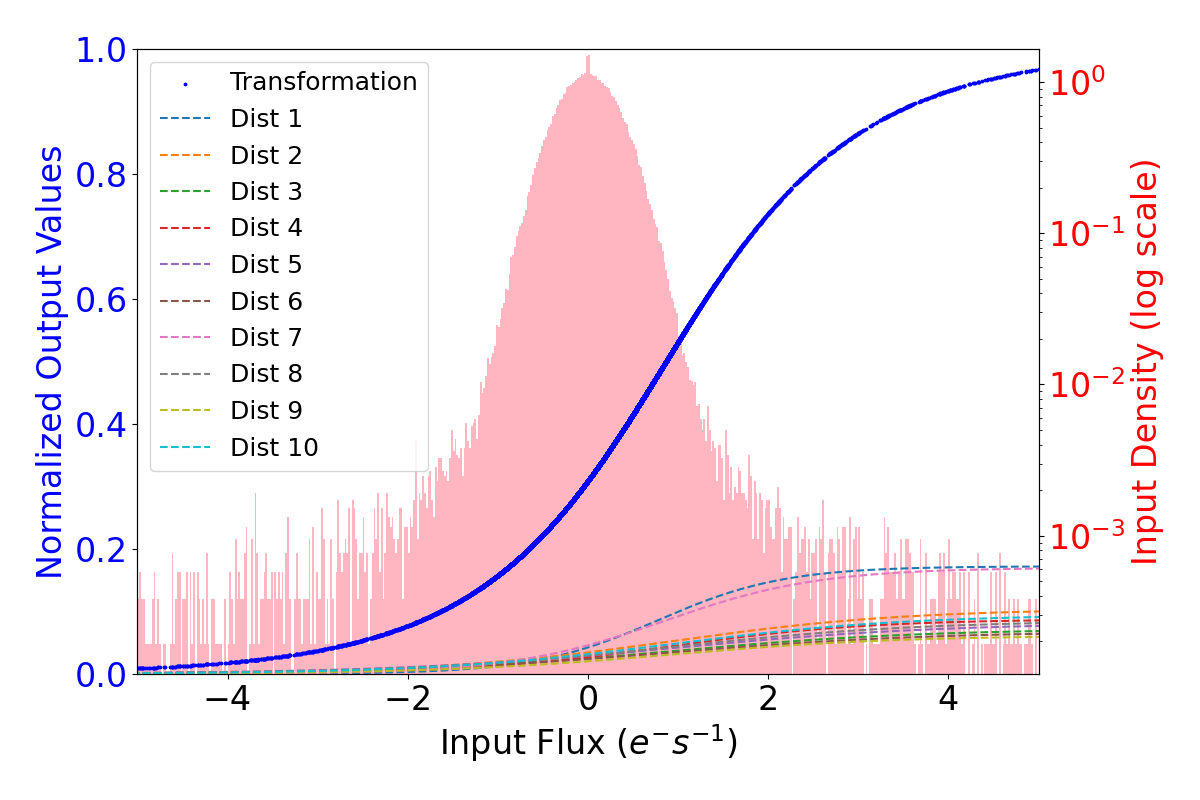}
    \caption{Example transformations from the Adaptive Normalization layer for the NT model ({\em top}) and the ST model ({\em bottom}). The histogram of the input data (median-subtracted flux of a randomly-selected sample) is shown in red, with density corresponding to the y-axis on the right side. Each of the ten weighted Logistic CDFs are shown as dashed lines, with the weighted sum shown in blue, all corresponding to the y-axis on the left side. The parameters of the Logistic CDFs are shown in Table \ref{tab:AN}.}
    \label{fig:AN}
\end{figure}

To demonstrate the functionality of this layer, we show the transformation of randomly selected data samples for both the NT and ST models in Figure \ref{fig:AN}, described in detail in the figure caption. The learned parameters for the Logistic CDFs comprising the transformation functions are shown in Table \ref{tab:AN}. Rather arbitrarily, we chose to have our transformation functions constructed by ten of the weighted Logistic CDFs. Greater or fewer CDFs can be chosen as desired, and we make no assertions as to an optimal number. In testing, however, we noticed that regardless of the number of CDFs selected for the transformation, the model tended to learn one or two as the dominant distributions, with progressively smaller contributions from others. In Figure \ref{fig:AN}, we see that the model learned one dominant CDF for the NT model and two almost equally dominant CDFs for the ST model.

\begin{table*}[ht!]
\centering
\begin{tabular}{|c|c|c|c|c|c|c|c|c|c|c|c|c|}
\hline
\multirow{2}{*}{Model} & \multirow{2}{*}{Param.} & \multicolumn{10}{c|}{Value} \\
\cline{3-12}
&  & 1 & 2 & 3 & 4 & 5 & 6 & 7 & 8 & 9 & 10 \\
\hline
\multirow{3}{*}{NT} & $\mu$ & 1.0156 & 1.0668 & 1.0730 & 1.0595 & 1.0243 & 1.0388 & 0.6247 & 1.0060 & 1.0225 & 1.0526 \\
\cline{2-12}
& $s$ & 1.1713 & 1.4406 & 1.4086 & 1.6195 & 1.1923 & 1.5156 & 0.4919 & 0.9800 & 1.6163 & 1.2180 \\
\cline{2-12}
& $w$ & 0.0432 & 0.1218 & 0.1354 & 0.1000 & 0.0626 & 0.0698 & 0.2207 & 0.0830 & 0.0389 & 0.1246 \\
\hline
\multirow{3}{*}{ST} & $\mu$ & 0.7804 & 0.9204 & 0.9154 & 0.9059 & 0.9201 & 0.9152 & 0.8493 & 0.9189 & 0.9091 & 0.9185 \\
\cline{2-12}
& $s$ & 0.6976 & 1.3089 & 1.4217 & 1.1051 & 1.5383 & 1.4734 & 0.8559 & 1.4186 & 1.2675 & 1.3183 \\
\cline{2-12}
& $w$ & 0.1722 & 0.1044 & 0.0725 & 0.0878 & 0.0823 & 0.0682 & 0.1699 & 0.0859 & 0.0616 & 0.0952 \\
\hline
\end{tabular}
\caption{Location ($\mu$), scale ($s$), and weight ($w$) parameters for the NT and ST models for each of the ten Logistic CDFs.}
\label{tab:AN}
\end{table*}

\begin{figure}
    \centering
    \includegraphics[width=.99\linewidth]{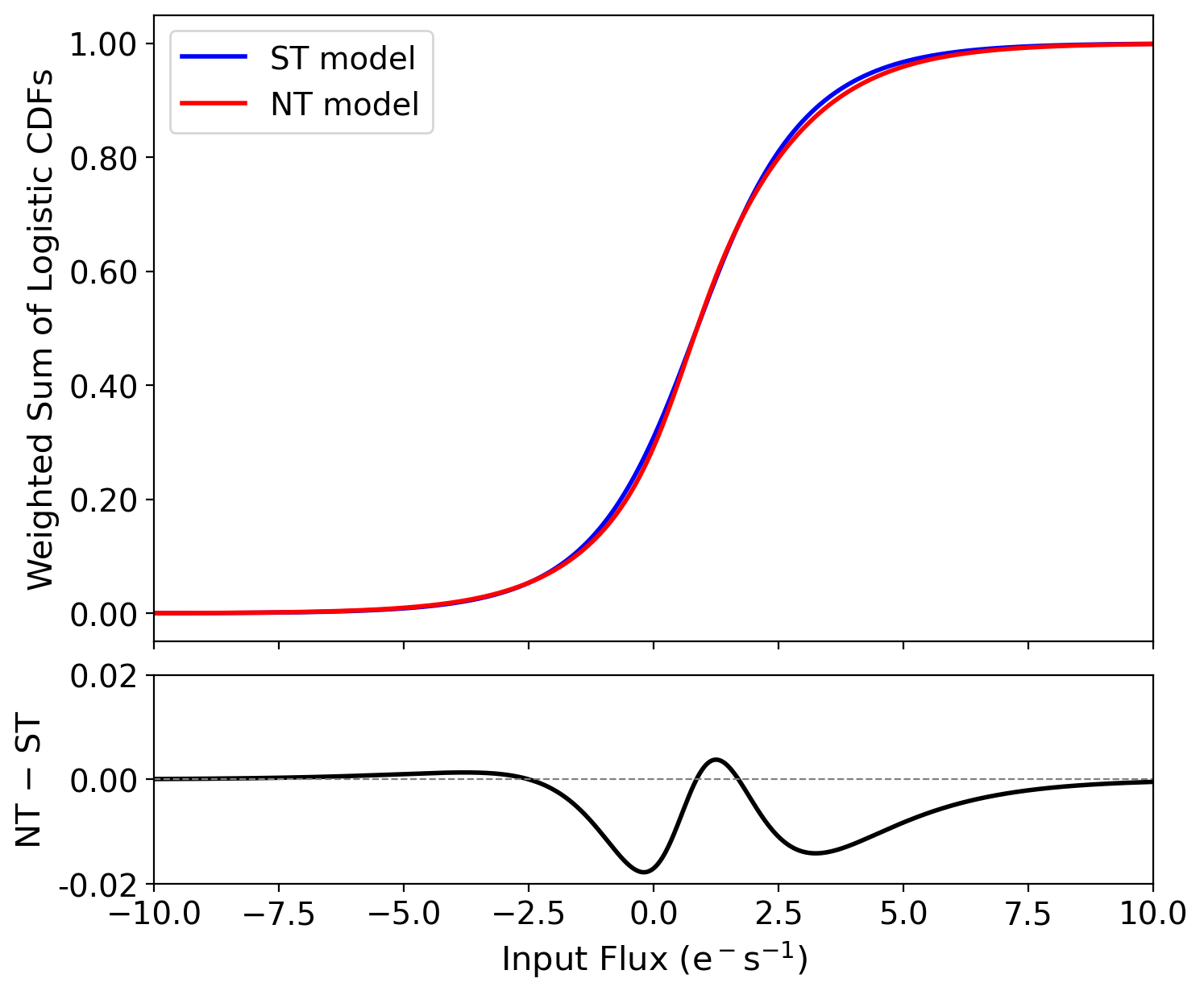}
    \caption{({\em top}) The learned transformation functions of the NT model (red) and the ST model (blue). ({\em bottom}) The difference between the transformation functions (NT-ST).}
    \label{fig:tf}
\end{figure}

We compare the learned transformation functions for the ST and NT models in the top panel of Figure \ref{fig:tf}, with the difference between them shown in the bottom panel. Despite the variety in the distribution parameters, the difference is rather slight, reaching a maximum absolute difference of $\sim$0.018 in the transformed output value. Interestingly, it can be seen from both this figure and the parameters themselves in Table \ref{tab:AN} that the learned locations, $\mu$, of the Logistic CDFs were all substantially positive. As discussed earlier in this section, positive $\mu$ was not a requirement of the model. We can consider the transformation function to be expanding the range of input flux near $\mu$, while contracting the range of input flux further away. In effect, it is allowing the neural network to intelligently select the ranges of the input flux which are most useful to determining the output. 

Given that the input flux is median-subtracted and the learned $\mu$ are all positive, this indicates that the model has learned a preference for greater detail in flux values slightly larger than the median. Logically, this is clearly intelligent behavior because an asteroid passing through a pixel will cause the flux from that pixel during the asteroid passage to be larger than the temporal median, barring other variability effects. This behavior inherently validates that our learned normalization technique performs as desired.

\section{Loss Function}
\label{sec:loss}
Semantic segmentation tasks have used a number of loss functions, each in consideration of a particular purpose. Generally, these loss functions focus on identifying particular pixels, regions, or boundaries between regions. Although it is beyond the scope of our effort to investigate each of them, we note that semantic segmentation loss functions, together with their various strengths and weaknesses, are nicely summarized by \citet{azad2023lossfunctionserasemantic}. Although many semantic segmentation problems have multiple classes in the same image data, we have only two classes: (i) pixels containing asteroids and (ii) pixels not containing asteroids, identified respectively in a binary mask as 1 and 0. The data set is heavily weighted toward the latter. 

\begin{figure}
    \centering
    \includegraphics[width=.99\linewidth]{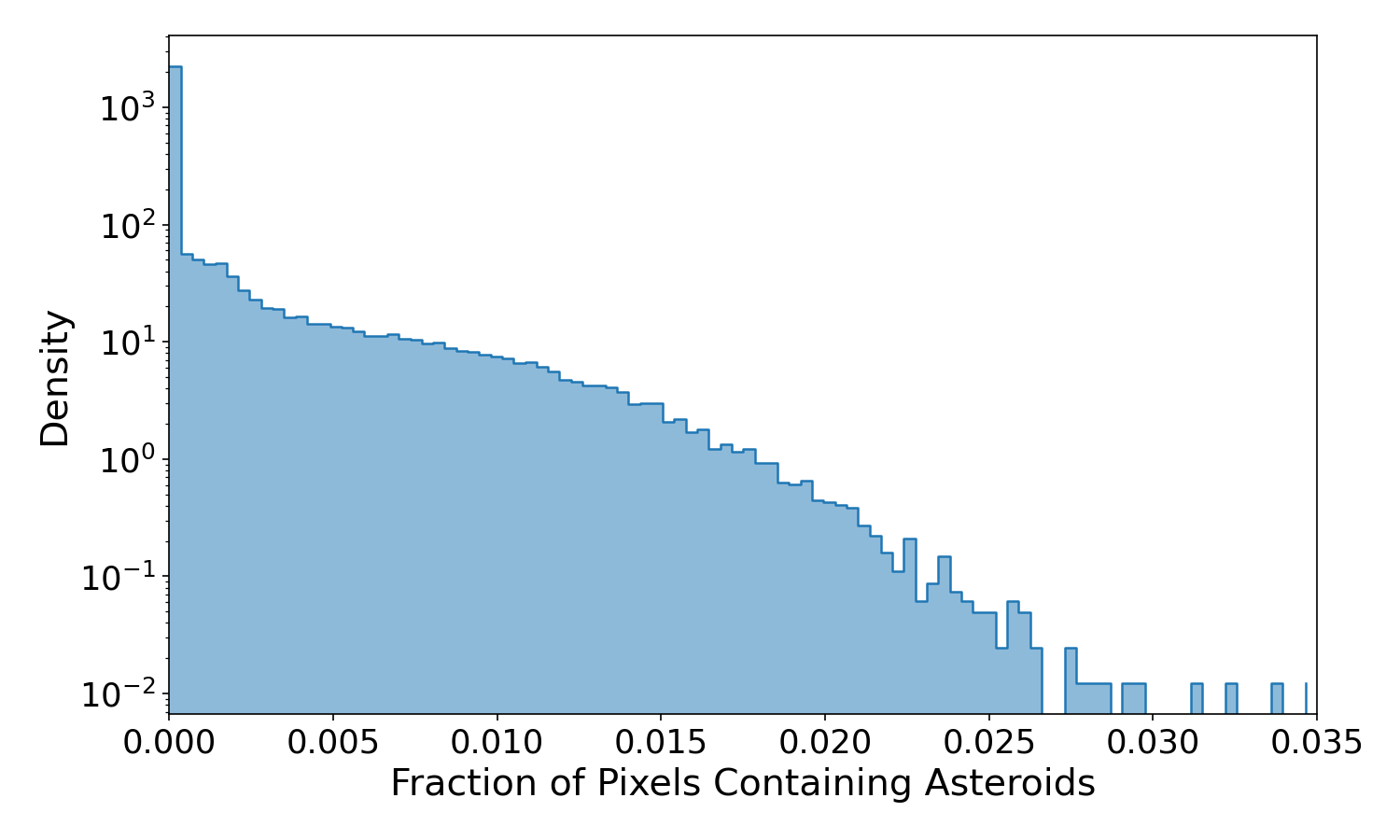}
    \caption{Histogram showing the fraction of pixels containing asteroids in all 64$\times$64$\times$64 data samples. Note that the y-axis is on a log scale and, as indicated by the large initial bin, $\sim$75.4\% of the data samples do not contain any asteroids.}
    \label{fig:frac}
\end{figure}

We initially started our effort with the intent of pixel-level segmentation given that we faced only a binary problem, correctable with inverse weighting of the pixels, but we quickly found that the imbalance between the classes was far too cumbersome to be overcome by an inverse weighting scheme. In fact, $\sim$75\% of our data samples contained no asteroids at all (see Figure \ref{fig:frac}, which shows a histogram of the fraction of pixels containing asteroids in each sample). In consideration of the loss function, the sparsity of the data simply implied that a pixel-based loss function was not feasible for the problem, but it also presented additional challenges for training the model, which we will discuss later in Section \ref{sec:batch}.

Due to the rather extreme class disparity, we needed to use a region-based loss function rather than a pixel-based loss function. We selected the generalized Dice loss function because it inherently manages class imbalance by effectively measuring the overlap between regions without consideration of class size. The Dice-S{\o}rensen coefficient was originally developed by \citet{dice1945measures} and \citet{sorensen1948method} as a measure of similarity between disparate samples, $A$ and $B$, given by

\begin{equation}
D=\frac{2|A\cap B|}{|A|+|B|} \ .
\end{equation}

The coefficient was originally proposed as a score of measuring overlap by \cite{1717643}, then first implemented as a loss function by \cite{2017arXiv170703237S} by subtracting the coefficient from unity. The Dice loss, $L_{D}$, is given by

\begin{equation}
L_{D}=1-\frac{2\sum_{n=1}^{N}p_{n}y_{n}}{\sum_{n=1}^{N}p^{2}_{n}+\sum_{n=1}^{N}y^{2}_{n}},
\label{eqn:dice_loss}
\end{equation}
where $N$ is the total number of pixels, $p$ is the true value of the pixel (0 or 1), $y$ is the predicted value of the pixel in the range [0,1]. Including the additional term from Eqn.~\ref{eqn:lv}, the model loss, $L$, becomes

\begin{multline}
L=L_{D}-L_{v}=\\
1-\frac{2\sum_{n=1}^{N}p_{n}y_{n}}{\sum_{n=1}^{N}p^{2}_{n}+\sum_{n=1}^{N}y^{2}_{n}}-w_{\mu}\frac{1}{1+e^{-\sigma^{2}(\mu)}} \ .
\label{eqn:loss}
\end{multline}
$L_{D}$ is limited in the range [0,1]. The range of the $L_{v}$ term is limited to  $L_{v} \leq w_{\mu}$ and is therefore user-defined. In our case, we used $w_{\mu}=0.01$, mildly encouraging distinct distributions from the Adaptive Normalization layer without distracting the model from the overall goal of minimizing the Dice loss.

\section{Model Training}
\label{sec:training}

\begin{figure*}
    \centering
    \includegraphics[width=.99\linewidth]{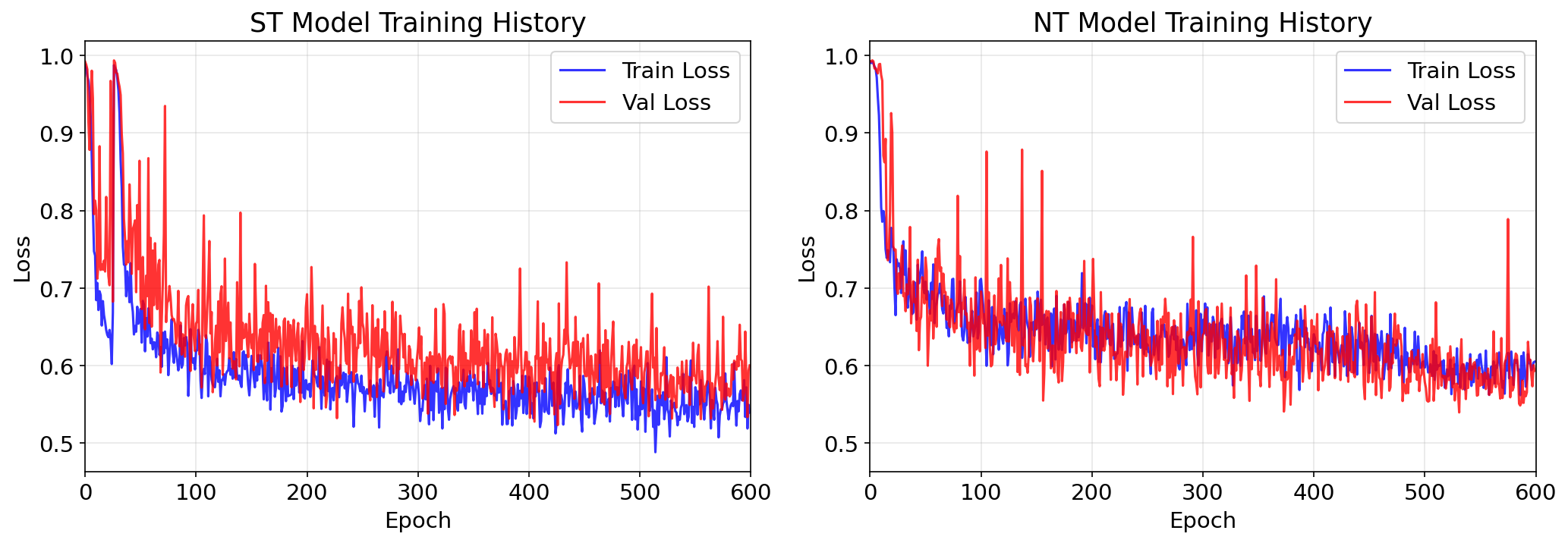}
    \caption{({\em left}) Training loss (blue) and validation loss (red) of the ST model over 600 epochs. ({\em right}) The same for the NT model.}
    \label{fig:training}
\end{figure*}

Training the neural network required several customizations. The data was not amenable to training in its raw form, nor was it conducive to a generalizable model. Here, we describe several challenges we encountered in the process of training and how they were overcome.

\subsection{Data Augmentation}

Referring back to our motivations discussed in Section \ref{sec:intro}, we wanted to avoid any dependency on asteroid trajectory in our method. By building a model equally capable of identifying asteroids moving in any direction through the FFIs, we could find asteroids which have been missed by traditional shift-and-stack searches dependent on high-probability direction and speed hypotheses. Since the capabilities of a ML model are, in large part, dictated by the nature of the training data, we needed to intervene in the training process by providing artificial data to the model.

Artificial, for our purposes, meant altering the configuration of the original data. Each 64$\times$64$\times$64 data cube provided as training or testing data to the model was randomly selected to be in one of 16 equally probable configurations: a 50\% chance of flipping the time axis, followed by one of eight possible rotations or flips on the positional axes. By training in this manner, the model became equally capable of identifying asteroids moving right-to-left as left-to-right, top-to-bottom as bottom-to-top, or any combination thereof.

We also note that shift-and-stack methods are generally limited to finding moving objects with tracks starting at a fixed initial time step, whether the first time step of the image stack or a  later time step set by padding. Our method suffers no such limitation as it is trained over a multitude of different asteroid tracks without any defined start or end point.

\subsection{Batch Construction}
\label{sec:batch}
We initially trained the neural network with a random selection of data cubes, but quickly found that the model was not capable of identifying asteroids, regardless of the batch size, structure, optimizer selection, or hyperparameter modifications. This was due to the fact that $>$75\% of cubes did not contain any asteroids, and the samples that did contain asteroids tended to be sparse. Since we were training to the objective of a binary mask, the neural network compensated for the lack of asteroids in the training data by having a tendency to output predictions containing all zeros.

In order to overcome this problem, we modified our batch construction to be careful to ensure that both our training and validation batches contained asteroids. We established an arbitrary minimum threshold of at least 100 pixels (0.04\%) of a data cube needing to contain an asteroid in order to be used for training or validation. 

The nature of the Dice loss function (previously discussed in Section \ref{sec:loss}) allows it to compensate for class imbalance in the data. However, in the case where the true mask consists of entirely zeros, the numerator of the second term of Equation \ref{eqn:dice_loss} is zero and thus the sample contributes no information of value to the loss function. Our arbitrary threshold ensured that each sample contributed to the loss, and therefore allowed for successful convergence of the model.  The NT and ST model training and validation losses over 600 epochs, demonstrating convergence, are shown in Figure \ref{fig:training}.  We attribute any differences between the model losses to differences in the data, as they are trained and validated on entirely different data sets.

\subsection{Data Completeness and Accuracy}
\label{sec:completeness}

We learned in the process of model development that the use of ML for asteroid detection is particularly problematic in that the ``truth'' data is fundamentally incomplete. In most image classification problems or even semantic segmentation problems, the truth is known. In our case, the truth is partial. By this we mean that, the dimmer our asteroid threshold, the greater the chance that an asteroid is not included in our truth data.

New asteroids are being found every day.\footnote{See the tallies updated daily by MPC on \url{https://www.minorplanetcenter.net/}}  Recent efforts \citep[e.g.][]{2021PASP..133a4503W,2024AJ....167..113N} have added thousands of new asteroids to existing databases. That the ``truth'' data on which we trained our model is incomplete is not in question, just the extent to which it is incomplete.

The impact this has on our model is effectively self-limitation. That is, the dimmer the object, the greater the chance we are training a pixel as a negative which is in reality a positive (i.e. an asteroid exists in a given pixel but has not yet been identified).

It is also possible that an asteroid in the database no longer exists on its recorded orbit or has diverged even slightly from the projected track indicated by the JPL Horizons ephemeris. In this case, we are rather severely impairing the training of the model, as we are both training as a positive what is in reality a negative and vice versa along the entire length of a track. This could amount to thousands of pixels for a single track which are being provided to the model with the wrong label.

The effect of both the fundamental incompleteness and inaccuracy of the data is that we are ``counter-training'' a non-trivial fraction of the data. That is, the neural network will be attempting to numerically justify why some pixels it understands as positives are actually negatives and vice versa. As such, the performance of our model, or any model attempting the same task, will be inherently limited by the completeness and accuracy of the asteroid data used for training. As can be expected (and will be discussed further in Section \ref{sec:results}), these limitations also hinder a proper quantitative evaluation of the results of the model. 

We also note that, although they are provided by JPL Horizons, we did not use comets in the training data.  Comets temporarily brighten pixels in their tails, which are not a characteristic of asteroids.  The tails can be rather large and cover a substantial pixel region, creating a confusing effect for the neural network.  As such, we chose to exclude comets from the training data by explicitly not training or validating with any data sample containing a comet, which is a flag provided by \texttt{tess-asteroid-ml}.  This of course does not preclude the model from finding comets, as they are moving objects just like asteroids, but it does prevent poor training of the model in the presence of comet tails.

\subsection{Magnitude Limitations}
\label{sec:magnitude_lim}

As with data completeness, inherent limits of the TESS data and our model cause the same counter-training. In Section \ref{sec:results}, we will discuss the detection magnitude limits of our model. Briefly, there exists a boundary at $V\approx21$ where any asteroid fainter than this value is effectively invisible to the model. We trained our model with asteroids down to a magnitude limit of $V\approx22$ in order to allow it to push the boundary as far as possible.

This was a calculated trade-off. If we did not train the model to find asteroids dimmer than $V\approx21$, then we effectively guarantee that the model will not be able to find asteroids dimmer than $V\approx21$. Training the model with asteroids to a limit of $V\approx22$ allows for this possibility, but also assumes the risk of counter-training, which in this case manifests as training pixels as positives that the model cannot perceive due to fundamental limitations of the data. We accepted the risk to up to this magnitude limit.

\section{Results}
\label{sec:results}

We performed inference on the entirety of our data set, using the NT model to evaluate every 64$\times$64$\times$64 sample of data from the Southern sectors and the ST model to evaluate every 64$\times$64$\times$64 sample of data from the Northern sectors. Referring back to Table \ref{tab:model_data}, this amounted to $\sim$\,490 million samples. Utilizing the PRISM GPU Cluster at the NASA Center for Climate Simulation (NCCS)\footnote{\url{https://www.nccs.nasa.gov/systems/ADAPT/Prism}}, we employed 4 GPUs per sector of TESS data, which required an average of $\sim$2.5 days to complete. Over 26 sectors, the full inference process required $\sim$6,000 GPU-hours.  This is a perfectly parallelizable process, so will scale directly with the number of GPUs.

The outputs of the model are 64$\times$64$\times$64 arrays in the range [0,1] indicating the likelihood of an asteroid in any given pixel as predicted by the model. Rather than evaluate the performance of the model on 64-timestep subsections of the data, we used one-timestep increments and averaged the model outputs on overlapping timesteps. For frame $n$ in an image stack with $N$ timesteps, the number of predictions averaged into the output, $p(n)$, is given by

$p(n) = \begin{cases}
n & \text{if } n \leq 64 \\
64 & \text{if } 64 < n < N - 63 \\
N - n & \text{if } n \geq N - 63. \\
\end{cases}$

While more computationally expensive (running inference on 64-timestep subsections would have required less than one day for the entire data set), this process made our predictions more robust and continuous, emphasizing true detections and minimizing spurious detections.

\begin{figure*}
    \centering
    \includegraphics[width=.99\textwidth]{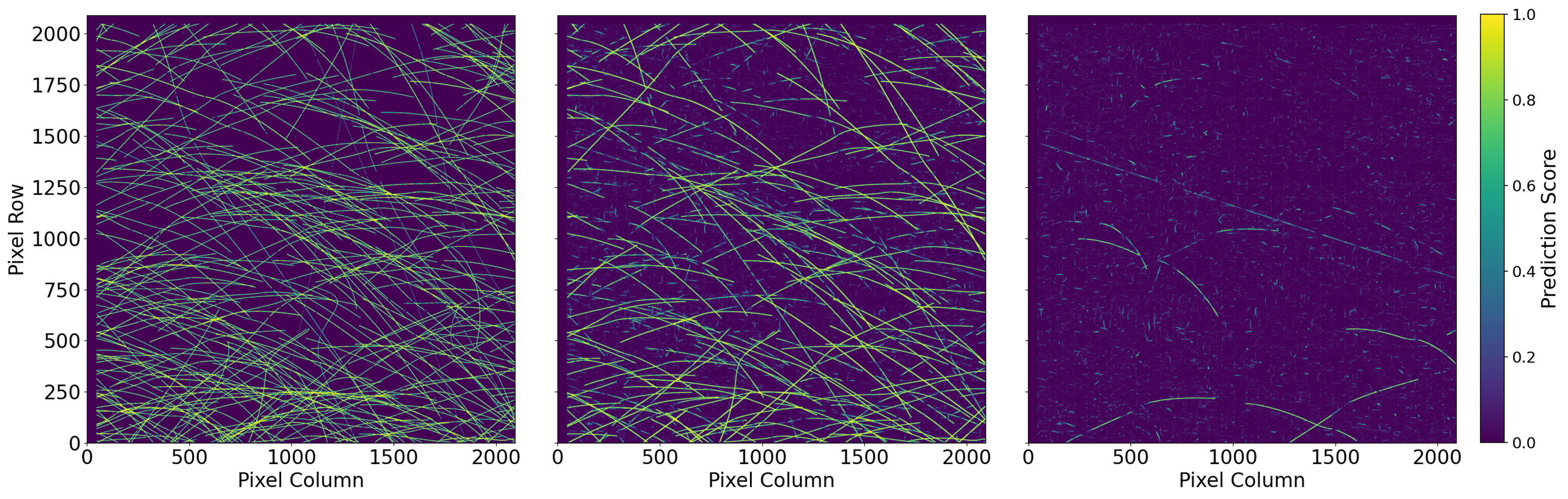}
    \caption{Temporal maximum aggregated results for Sector 3, Camera 2, CCD 4. ({\em left}) JPL tracks of known asteroids, ({\em middle}) our model prediction, and ({\em right}) the difference between the two (indicating tracks not present in the training data). The JPL Horizons track plots are binary (0 for no asteroid, 1 for asteroid), whereas our model outputs are in the range [0,1].  The longer tracks in the right panel are very clear detections over multiple data cubes, which are certainly real tracks.  The short-length detections in the same panel are either (i) detections of asteroids not present in the training data, (ii) residuals between the detected tracks and the JPL Horizons tracks, or (iii) false positives.  In follow-on processing, we do not consider a track to be valid unless it spans more than one data cube (i.e.~$>$ 64 pixels).}
    \label{fig:example_output}
\end{figure*}

With these aggregated model outputs, we were able to compare our results directly to the projected JPL Horizons tracks. We provide an example of the aggregated outputs for our model for Sector 3, Camera 2, CCD 4 in Figure \ref{fig:example_output}. The left panel contains the JPL Horizons tracks as projected onto the TESS data, the middle panel shows the temporal maximum value of our aggregated model outputs, and the right panel shows the difference of the JPL Horizons tracks subtracted from the model outputs in order to show tracks found by our model that were not present in the training data. While our model certainly does not find all the tracks from JPL Horizons (to be discussed in Section \ref{sec:completeness}), especially of dimmer asteroids, it does indeed find some that are not present in the training data.

\begin{figure*}
    \centering
    \includegraphics[width=.99\textwidth]{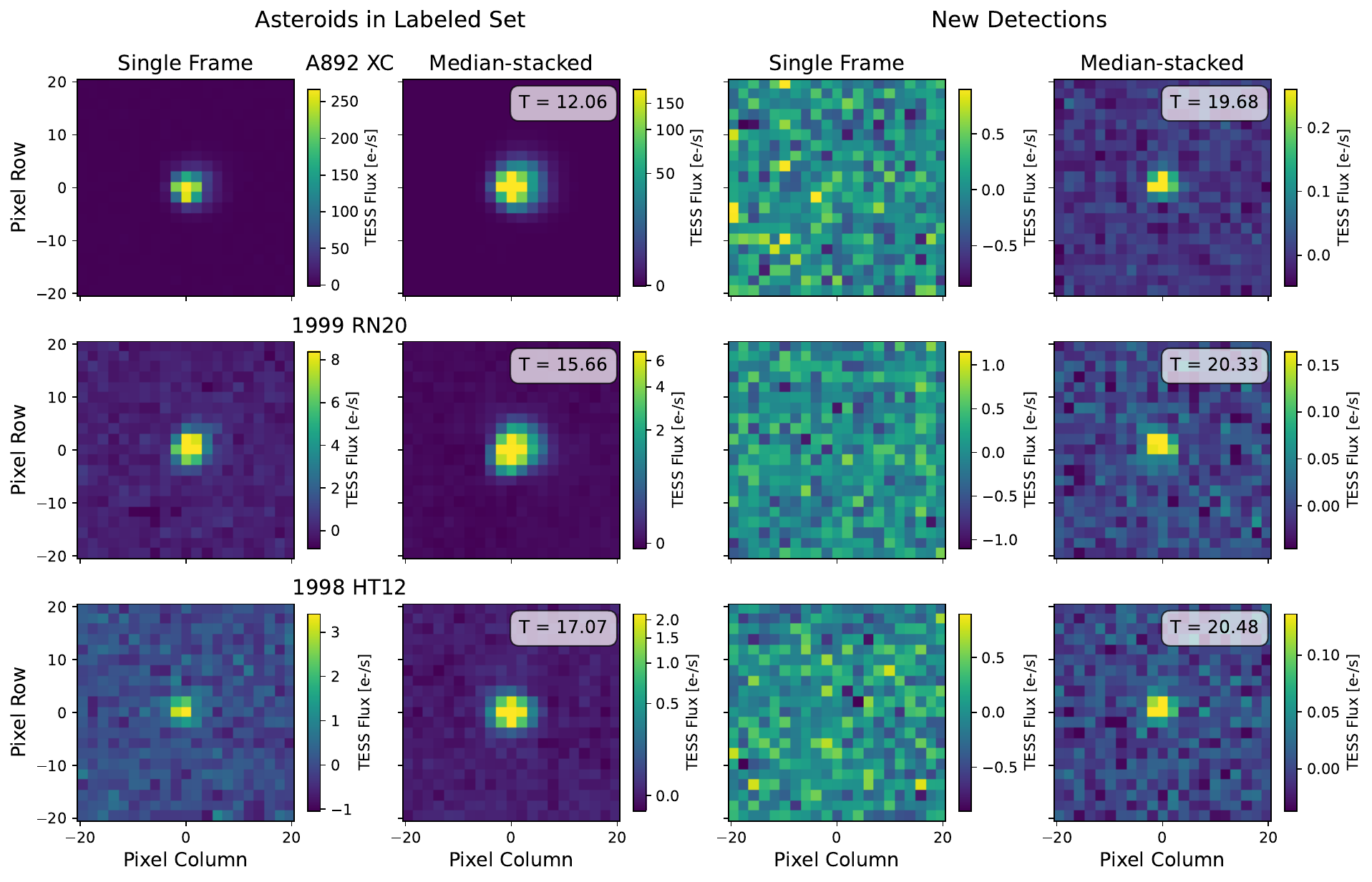}
    \caption{Examples of six detected sources in Sector 3, Camera 2, CCD 4 (from Figure \ref{fig:example_output}). The left two columns show asteroids in the labeled set (names in panel titles) and recovered by our model. The right two columns show new detections (not in the labeled set). A comparison between single frame and median-stacked images are given for each source to highlight the signal at different TESS magnitudes (top right legends). 
    }
    \label{fig:stacked_tracks}
\end{figure*}

To inspect the veracity of these detections we computed their locations at each frame by averaging pixel rows and columns using the detection probabilities as weights. We fit b-spline third order polynomials to the detections to obtain the source tracks. Figure \ref{fig:stacked_tracks} shows six example detections, including three known asteroids recovered from the labeled set and three new detections. These TESS FFI cutouts were created using \texttt{tess-asteroids}\footnote{\url{https://github.com/altuson/tess-asteroids}} -- an open-source python package to create image cutouts and lightcurves for asteroids observed by TESS \citep{tuson_2025_16332750}. 

None of the new objects can be visually detected in the single-frame images, but after median-stacking the images using the detected tracks we obtain detections with $\text{SNR}>8$ and TESS magnitudes between $19.7 < T < 20.5$. The three examples of known asteroids are included to exemplify how brighter asteroids are detectable even in single-frame images.
This indicates that our model effectively learns the shift-and-stack method to detect sources fainter than the single-frame detection limit of the data. While shift-and-stack requires users to input a finite number of velocity vectors on which to stack, our model implicitly stacks in any and all directions through the filters of the neural network. This means that while shift-and-stack may be biased towards or under represent certain directions, our model is unbiased, comprehensive, and efficient.
Although out of scope for this paper, additional vetting is currently underway to confirm or deny our ``new'' tracks as genuine newly discovered tracks, which we plan to elaborate upon in future work.

\subsection{Quantitative Comparison}
\label{sec:quant}
In the previous section, we discussed the challenges posed by the inexact nature of our data in the process of training. Here, we face the same challenges with regard to quantifying the performance of our model. We re-emphasize that there is no guarantee that our ``truth'' data is actually true. Hereafter, we will refer to the projections of the JPL Horizons tracks as truth data (without the quotes), but we remind the reader to understand the nature of this data as we previously described.

\begin{figure}
    \centering
    \includegraphics[width=.98\columnwidth]{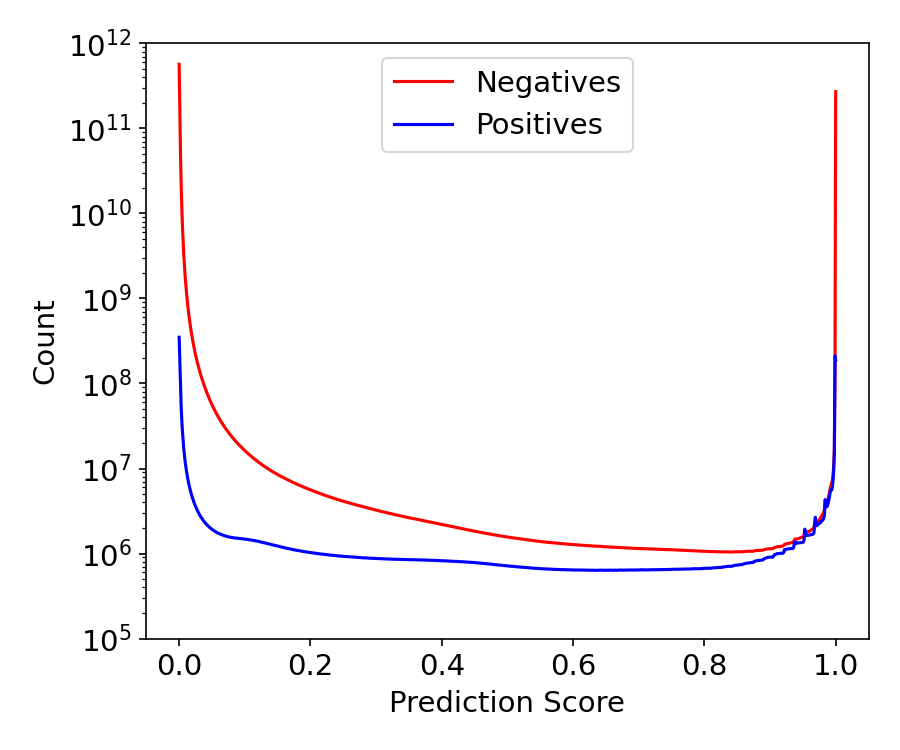}
    \caption{Distribution of per-pixel prediction scores according to the truth data positives ({\em blue}) and negatives ({\em red}). The counts represent binned data at a width of 0.001.  Note that, although the domain of the distributions is [0,1], we extend the x-axis on both ends for visual purposes as much of the distributions are close to the limits.}
    \label{fig:pndist}
\end{figure}

First, we will discuss the distribution of our predictions, combined for all of TESS Sectors 1-26. Figure \ref{fig:pndist} shows the per-pixel distributions (binned at width 0.001) of our prediction scores corresponding to positive and negative pixels from the truth data. Note that only $0.2\%$ of the truth data are positives, hence the disparity in the number counts. Both the positive and negative distributions have a similar shape in that the model predictions are either strongly positive (1) or strongly negative (0), with relatively few predictions in the intermediate range. The distribution of negatives is more weighted toward lower predictions, but still has a substantial number of strong positives. Indeed, there are more true negatives predicted as positives than there are true positives predicted as positives. We will be specific here about the prediction counts in the extreme ends, because they will heavily influence the analysis moving forward. In the prediction range [0, 0.001] there were $\sim$$5.7\times10^{11}$ negatives and $\sim$$3.5\times10^{8}$ positives, while in the prediction range [0.999, 1] there were $\sim$$2.7\times10^{11}$ negatives and $\sim$$1.9\times10^{8}$ positives. An ideal model evaluation, of course, would have the negatives strongly weighted toward zero and the positives strongly weighted toward unity. In our case, the model is nearly certain that many pixels that are negatives in the truth data are actually positives and vice versa. In fact, in each case, the model is near-certain that approximately half of the total number of pixels are contrary to their labels. This is the initial indication of the difficulty in quantifying our results given the nature of the truth data. These distributions will heavily influence subsequent results.

Because of the data imbalance (again, only 0.2\% of the pixels are positives in the truth data), model accuracy is not a valid metric. As such, the results are better demonstrated using precision-recall (PR) curves or receiver operating characteristic (ROC) curves. We will show both for clarity.

Both PR curves and ROC curves evaluate the results at prediction thresholds, such that given labeled positives $P$ and labeled negatives $N$, for prediction score $q$ in the range $[0,1]$, 

\begin{equation}
    \text{TP}=\sum_{P: P_i > q} P_i
\end{equation}
\begin{equation}
    \text{FP}=\sum_{P: P_i < q} P_i
\end{equation}
\begin{equation}
    \text{TN}=\sum_{N: N_i < q} N_i
\end{equation}
and
\begin{equation}
    \text{FN}=\sum_{N: N_i > q} N_i.
\end{equation}
Precision and recall are defined as
\begin{equation}
    \text{Precision} = \frac{\text{TP}}{\text{TP}+\text{FP}}
\label{eqn:precision}
\end{equation}
and
\begin{equation}
    \text{Recall} = \frac{\text{TP}}{\text{TP}+\text{FN}}
\end{equation}

It is important to emphasize that, with precision, we are mixing a quantity derived from the labeled positives (TP) with a quantity derived from the labeled negatives (FP), which will lead to lower precision values for brighter asteroids since there is a smaller quantity. We will show precision and recall for all asteroids less than a given magnitude, but we will compensate for this mismatch by inverse weighting the FP term of precision by the number of positive pixels for each magnitude level divided by the number for $V\leq22$. The full model performance should be considered as the performance for $V\leq22$.

\begin{figure*}
    \centering
    \includegraphics[width=.99\textwidth]{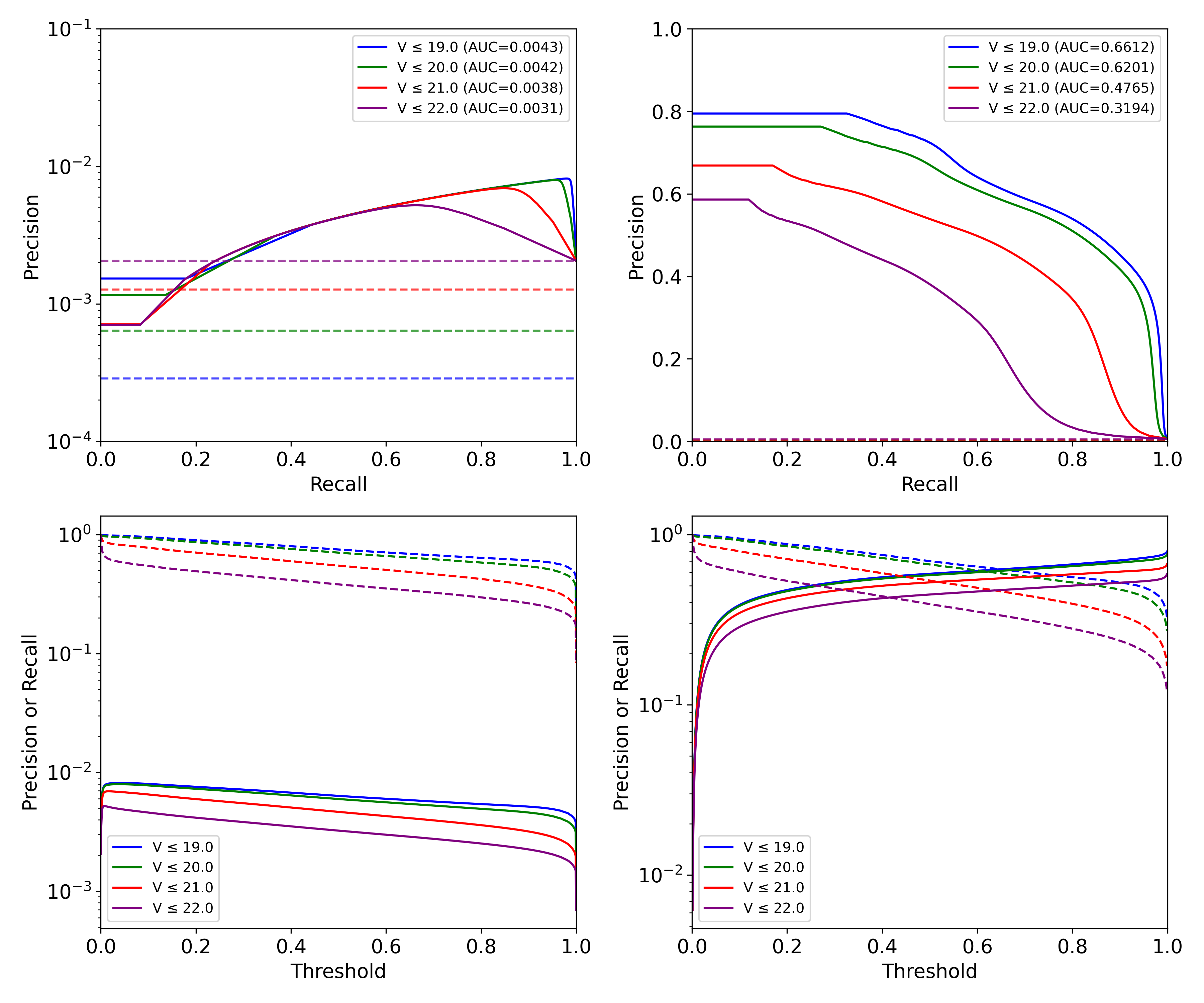}
    \caption{({\it top left}) Solid lines show the PR curve for the labeled positives at $V<19$ (blue), $V<20$ (orange), $V<21$ (green), and $V<22$ (red). The dashed horizontal lines show the value of precision at which the model has no discriminative ability for each magnitude threshold (i.e. the outputs are equivalent to randomly guessing). ({\it top right}) Same as top left, but without threshold bins $>$0.999 or $<$0.001. ({\it bottom left}) Threshold vs precision (solid lines) and recall (dashed lines). ({\it bottom right}) Same as bottom left, but without threshold bins $>$0.999 or $<$0.001.}
    \label{fig:pr_all_data}
\end{figure*}

In Figure \ref{fig:pr_all_data}, we show the PR curve of the full results in the top left panel, stratified by magnitude ($V<19$: blue, $V<20$: green, $V<21$: red, $V<22$: purple). The solid lines are the PR curves, and the dashed lines are the baseline, i.e.~the performance of a ``junk'' model with no discriminative ability. The precision baseline is the fraction of all the data which are the minority class, which in our case are the positives. This number is different for each magnitude threshold, with a maximum at the previously mentioned value of $0.2\%$ for $V\leq22$. A model with a PR curve above the baseline demonstrates validity, otherwise not.  

If we accept the PR curve from the upper left panel of Figure \ref{fig:pr_all_data} as our quantitative model performance, then our outlook would be rather grim. The PR curve can be evaluated using the area under the curve (AUC) as a model performance metric.  The AUCs (noted in the legend) are tiny, and the portion of the curve below the junk model baseline is also troubling. However, we need to consider our results in the context discussed in Section \ref{sec:training}. Specifically, there is no guarantee that our labeled positives are actually positive or that our labeled negatives are actually negative. Therefore, we re-evaluate our results with a simple assumption: when our model is very certain of a prediction, regardless of the truth data label, it is correct. To this end, we will be extremely conservative in assuming that model predictions greater than 0.999 are, in fact, positive, and model predictions less than 0.001 are, in fact, negative. Rather than reassigning them as true positives and true negatives, if we simply ignore the extreme bins on either end, we produce the PR curve shown in the upper right panel of Figure \ref{fig:pr_all_data}. The AUCs (noted in the legend) are much improved. The performance of the full model, since all magnitudes comprise a single class, is best understood as the AUC for $V\leq22$, which, at $\sim$0.32, is quite promising. The PR AUC needs to be considered in reference to the baseline, which, for our highly imbalanced data, is at Precision = 0.002, reflecting the 0.2\% of pixels containing asteroids.  This value would be the PR AUC of a model with no discriminative ability.  With a PR AUC of $\sim$0.32, our model outperforms a random selection model by more than 150$\times$.  We also suggest that the PR AUC is actually much better than we show here, as the right panels of Figure \ref{fig:pr_all_data} only exclude the extreme bins as an attempt to mitigate the effects of mislabeling in our evaluation.  The lower right panel again shows the precision (solid lines) and recall (dashed lines) against the threshold prediction value. Compared to the same plot in the lower left panel (which contains the extreme threshold bins) we can easily the improvement in precision which drives the improvement in the PR curve.

The ROC curve is a measure of the False Positive Rate (FPR) versus the True Positive Rate (TPR). These are defined as
\begin{equation}
    \text{TPR} = \frac{\text{TP}}{\text{TP}+\text{FN}}
\end{equation}
and
\begin{equation}
    \text{FPR} = \frac{\text{FP}}{\text{FP}+\text{TN}}.
\end{equation}
Note that TPR is identical to recall, just named differently in this context. We emphasize that, in the case of imbalanced data, the ROC curve is a much less reliable metric for model evaluation than the PR curve due to the FPR term.  In our dataset, the number of TN will always be dramatically higher than FP, making the FPR appear artificially low even in the case of a disproportionately large number of FP.  With this caveat, we note that we show the ROC curve for completeness, but the PR curve is more informative.

\begin{figure*}
    \centering
    \includegraphics[width=.99\textwidth]{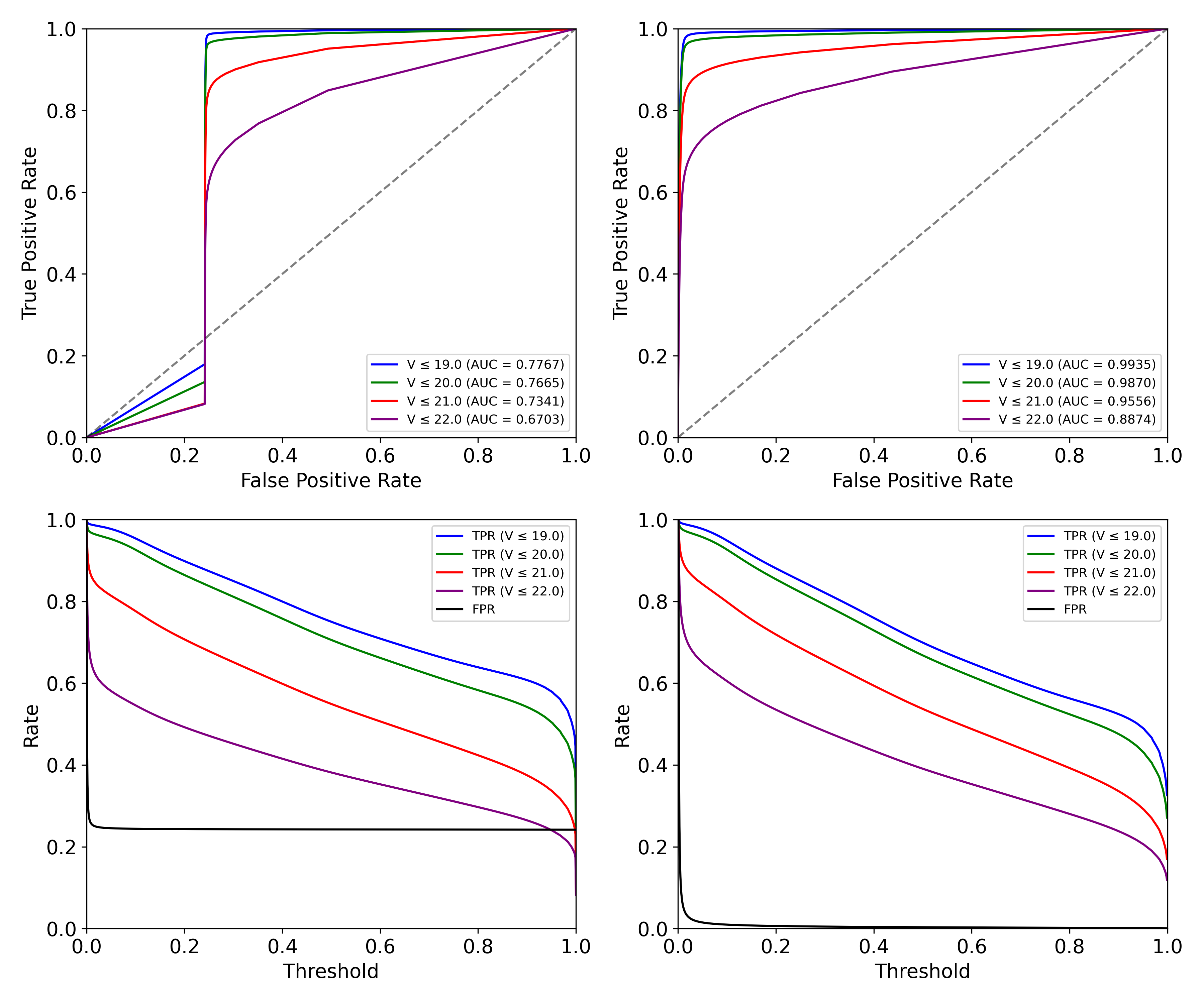}
    \caption{({\it top left}) ROC curve for the labeled positives at $V<19$ (blue), $V<20$ (orange), $V<21$ (green), and $V<22$ (red). The black dashed diagonal line is at FPR = TPR for reference. ({\it top right}) Same as top left, but without threshold bins $>$0.999 or $<$0.001. ({\it bottom left}) Threshold vs TPR and FPR rates, with colors the same as above for TPR. FPR is shown in black. ({\it bottom right}) Same as bottom left, but without threshold bins $>$0.999 or $<$0.001.}
    \label{fig:roc_all_data}
\end{figure*}

The ROC curve shows a continuous relationship between the TPR and FPR at any given prediction score threshold and, like the PR curve, can be evaluated using the AUC as a model performance metric. In Figure \ref{fig:roc_all_data} we show the ROC curves accumulated for all asteroids less than a given magnitude ($V<19$: blue, $V<20$: green, $V<21$: red, $V<22$: purple). The black dashed diagonal line indicates FPR = TPR, so for any point beneath this line, FPR $>$ TPR therefore the model has no predictive value. Consider first the upper left panel. The ROC curve shape is notably strange with the apparent discontinuity. So, again, we return to the uncertain nature of the truth data. To examine this behavior, in the lower left panel we show the TPRs for each magnitude (in the same colors as the upper panel) and the FPR (which is the same for all magnitudes, in black). The TPRs are as expected, decreasing gradually with increasing threshold. The FPR, however, quickly approaches a rather large limiting value. This value is driven by the ``U'' shape of the distribution of negatives shown in Figure \ref{fig:pndist}. Specifically, the far right of the distribution where the model is near-certain that a substantial portion of the data labeled as negatives are actually positives.

We followed the same process as with the PR curves and re-evaluated the ROC curves, removing consideration of the bins for either extreme due to the inherent ambiguity of the data. This change resulted in the much-improved ROC curve shown in the upper right panel of Figure \ref{fig:roc_all_data}. The AUCs (noted in the legend) are substantially higher, with the AUCs for $V\leq19$ and $V\leq20$ near unity. In the lower right panel of Figure \ref{fig:roc_all_data}, the TPRs are very similar to the lower left panel, with the primary difference being in the reduction of the FPR.

Additional methods are available for quantitative evaluation of model performance on imbalanced data, e.g. a confusion matrix or sensitivity-specificity, among others. However, each of these would inevitably lead to the same conclusion: The model performance is extremely poor if the truth data is considered to be absolute, but quite good when assuming that the model frequently knows better than the truth data, especially in the case of near-certain ``false'' positives. As such, with substantial barriers to quantitative evaluation of our model performance, we proceed to measure completeness.

\subsection{Completeness}
\label{sec:completeness}
\begin{figure}
    \centering
    \includegraphics[width=1.0\columnwidth]{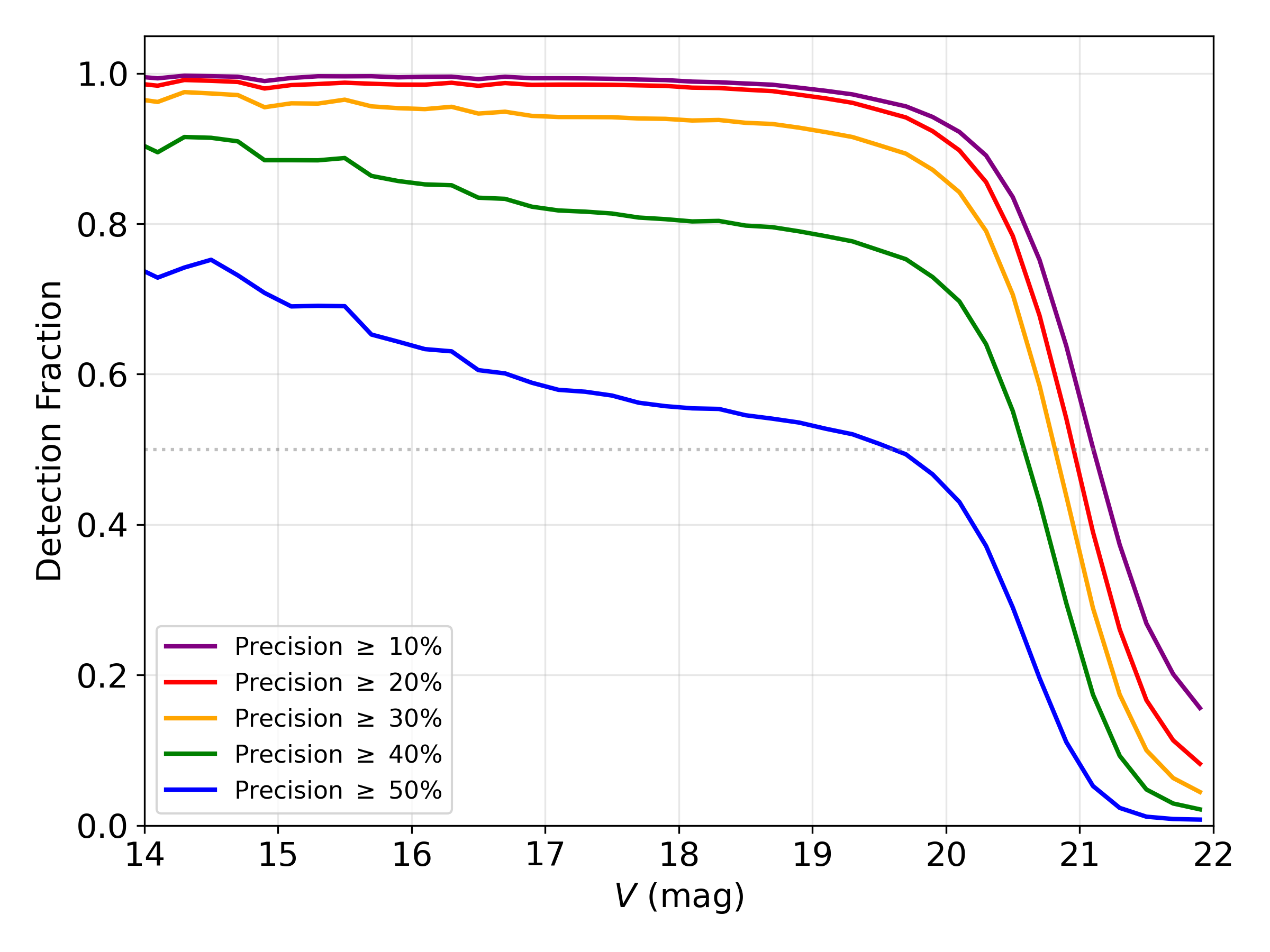}
    \caption{Detection fraction (or completeness) as a function of asteroid visual 
    magnitude at fixed pixel-level precision levels for all labeled asteroids ($V \leq 22$). Each curve corresponds to the prediction 
    score threshold required to achieve the indicated precision (given in Table \ref{tab:completeness_precision}), where precision represents the fraction of all model-selected positive pixels that correspond to labeled positives. The horizontal gray dotted line marks 50\% detection efficiency.}
    \label{fig:det_frac}
\end{figure}

In order to evaluate completeness of our model, we collected the per-pixel prediction scores of each pixel containing a given asteroid. For us to quantify our model detection threshold, the question arises of what prediction score comprises a detection. A decrease in the threshold will find more asteroids, but will also increase spurious detections. An increase in the threshold will decrease spurious detections, but will detect fewer asteroids. To address the ambiguity in threshold selection, we report completeness in terms of fixed precision values of 10\%, 20\%, 30\%, 40\%, and 50\% in Figure \ref{fig:det_frac}.

The remarkable feature of this plot is the ``wall'' between $V=20$ and $V=22$ indicating the limit at which asteroids are capable of being detected in TESS data. The high amplitude, spatially varying TESS scattered background light causes an increase in photon noise in TESS images, which may contribute to this ``wall'' where detections are no longer possible. In this work we do not investigate whether particular times where TESS scattered light is lower in amplitude affect the detection magnitude limit. Interpolated 50\% detection thresholds for each of these curves are given in Table \ref{tab:completeness_precision} along with total pixel counts for TP and FP. We emphasize that the counts represent individual pixels, not detections or tracks.  A single asteroid track typically spans many thousands of pixels across multiple frames and, similarly, a single artifact or systematic features can contribute many contiguous false positive pixels.  We also emphasize, again, that labeled FP are quite possibly TP.  The pixel-level precision demonstrates that the model assigns high confidence preferentially to pixels containing asteroids.  The maximum achievable pixel-level precision within the evaluated threshold range (0.001 - 0.999) is $\sim$59\%.  Higher precision is likely achievable, but cannot be reliably evaluated due to the label ambiguity.

\begin{deluxetable}{ccrrc}
\tablecaption{Interpolated $V_{50\%}$ detection performance at fixed pixel-level precision for all labeled asteroids ($V \leq 22$), with TP and FP pixel counts reported at each threshold. 
\label{tab:completeness_precision}}
\tablehead{
\colhead{Precision} & \colhead{Threshold\tablenotemark{a}} & \colhead{TP} & \colhead{FP\tablenotemark{b}} & \colhead{$V_{50\%}$ (mag)}
}
\startdata
10\% & 0.015 & 1,265,869,342 & 10,850,539,169 & 21.10 \\
20\% & 0.043 & 1,167,872,438 & 4,628,391,456 & 20.95 \\
30\% & 0.115 & 1,050,714,262 & 2,448,922,389 & 20.82 \\
40\% & 0.319 & 838,564,320 & 1,256,446,578 & 20.58 \\
50\% & 0.792 & 502,373,981 & 502,244,532 & 19.61
\enddata
\tablenotetext{a}{The model prediction score threshold necessary to achieve the given precision level.}
\tablenotetext{b}{Many FP likely correspond to genuine asteroids not yet present in the JPL Horizons catalog or asteroids that have diverged from the JPL Horizons ephemeris, making the reported precision values lower bounds on the true precision.}
\end{deluxetable}

With the caveat that none of the models are directly comparable, examine our Figure \ref{fig:det_frac} in the context of Figure 13 of \citet{2021PASP..133a4503W}, Figure 5 of \citet{2024AJ....167..113N}, or Figure 6 of \citet{2025AJ....170..187Z}. We do not claim better or worse performance than any of these efforts, as the models are evaluated with substantially different methodologies. Both \citet{2021PASP..133a4503W} and \citet{2024AJ....167..113N} fit their discovered tracks and compare the ephemerides directly to known tracks, which is superior to our pixel-level comparison due to the errors involved in our method, discussed previously. Although pixel-level comparison is suitable for our purposes in this initial study, we would plan to emulate \citet{2021PASP..133a4503W} and \citet{2024AJ....167..113N} in future work with our method.

\subsection{Velocity and Inclination Considerations}

Referring back to our intent with this model development, discussed in Section \ref{sec:intro}, we wanted to create a ML model for asteroid detection free of the velocity and inclination assumptions required for shift-and-stack models. The most interesting asteroid orbits will be those that are substantially different from known asteroids, hence they will occupy a different part of the parameter space compared to where most asteroids are currently found. To that end, we provide an evaluation of our model's performance relative to these parameters. 

\begin{figure*}
    \centering
    \includegraphics[width=.98\textwidth]{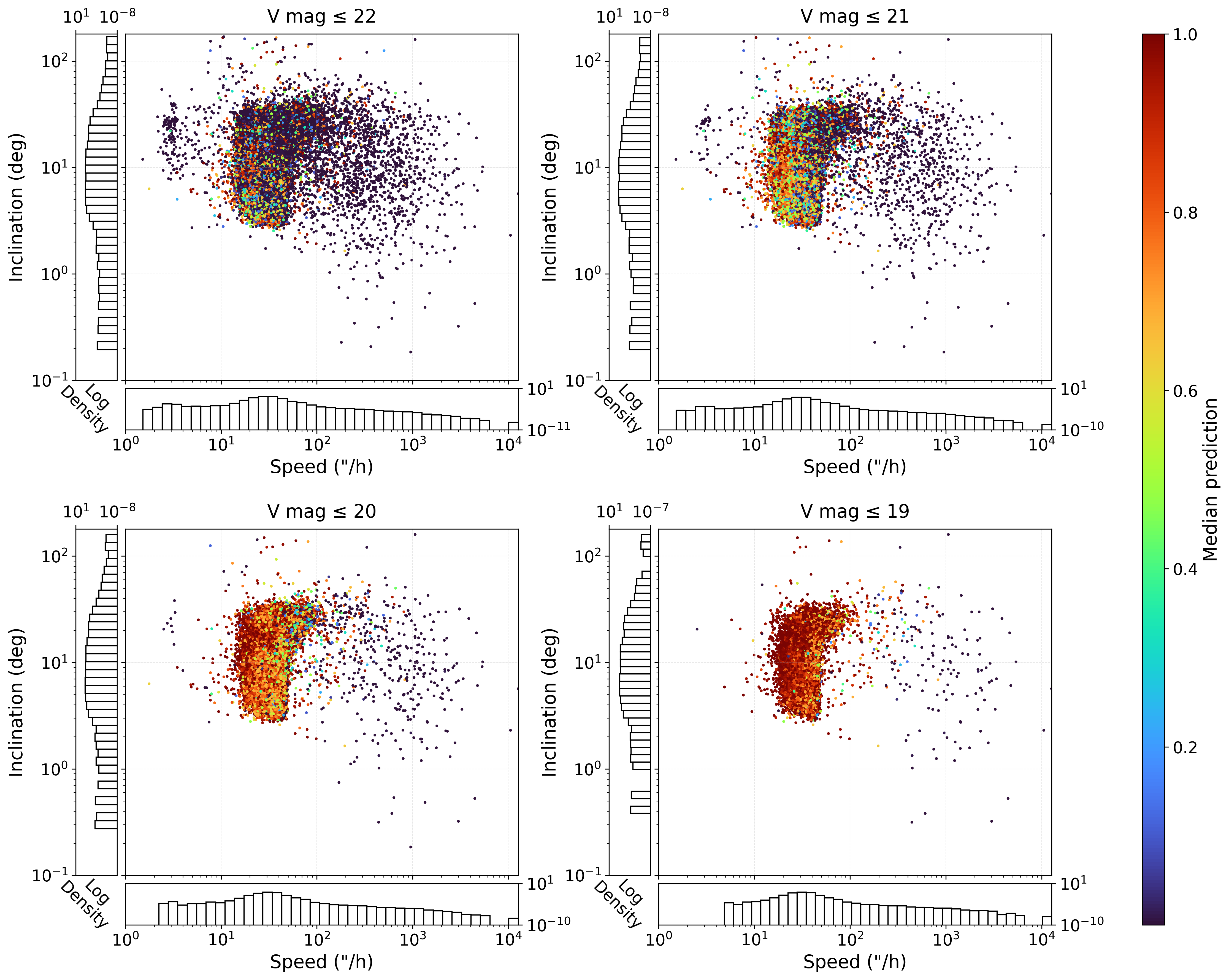}
    \caption{Asteroid absolute inclination (deg) versus speed (deg/hr) for magnitude thresholds 22 (upper left), 21 (upper right), 20 (lower left), and 19 (lower left). The color indicates the median prediction given by the colorbar on the right.}
    \label{fig:inc_vs_speed}
\end{figure*}

Figure \ref{fig:inc_vs_speed} shows a scatter plot of the inclination vs speed for each known asteroid, along with our median prediction for the pixels containing any given asteroid shown by the color. The low predictions dominating the plot of threshold $V\leq22$ (upper left panel) are reflective of the limiting magnitude shown in Figure \ref{fig:det_frac}. As the magnitude threshold is reduced, the predictions generally improve as the color turns more red (i.e. toward unity). Examining the $V\leq19$ plot (lower right panel), we can see that there does not appear to be any obvious trend in prediction score in the main distribution along the inclination axis. 

We do, however, note that very high speed asteroids ($\gtrsim$300\arcsec/hr) have a tendency for low prediction scores. The cause of the model's difficulty with extremely fast asteroids is easily understood with `back of the envelope' calculations. A single TESS pixel has an edge length of 21\arcsec, meaning an asteroid moving at 300\arcsec/hr will cover $\sim$14 pixels/hr. At this rate, and with a 30-minute effective exposure time, a single frame will show a blurred image about 7 pixels long. In the even more extreme case of an asteroid moving 10,000\arcsec/hr, the object will cover $\sim$240 pixels in a single cadence. Longer blurs also dilute the flux in any given pixel, lowering the SNR of the signal. These blurs are not entirely undetectable, but certainly not ideal for a model trained to find differences in nearby pixels by timestep. Rather than a weakness of the algorithm, we consider this a limitation of the training data.  Since we trained the model on real data, it learned to optimize for the real distribution of asteroid speeds.  As such, it did not learn well how to accommodate the long blurs of rare fast asteroids.  In retrospect, synthetic injections of fast asteroids into the training data would likely improve the model's ability to detect them.  

Given this understanding, we qualitatively evaluated the remainder of the data, where we would describe the performance of the algorithm on the speed axis to have a slight downward trend with increasing speed, as we note the transition from dark red to light red as speed increases through the main distribution in the $V\leq19$ plot. We again consider this trend to be a consequence of the asteroid speed distribution in the training data.

In consideration of this analysis, we assert that the performance of the model is largely invariant to inclination, while only slightly sensitive to increased speed within the limitations of the instrument. As such, we consider our objective accomplished. We also expect that in subsequent years of TESS data with shorter cadences (or in other missions with shorter cadences), that faster asteroids will be easier to detect with our algorithm due to shorter length blurs in single cadences.  Again, we also suggest that fast asteroids be injected into the training data in future applications in order to train the model to a more uniform distribution of asteroid speeds.

\section{Discussion}
\label{sec:discussion}

Our deep learning model, the W-Net, demonstrated clear success in detecting asteroids in TESS data, as shown by Figures \ref{fig:example_output} and \ref{fig:stacked_tracks}.  The right panels of Figures \ref{fig:roc_all_data} and \ref{fig:pr_all_data} demonstrate this quantitatively as well with ROC and PR metrics, despite the inherent difficulties of the data.  Our detection fraction, shown in Figure \ref{fig:det_frac}, is a similar result to the leading algorithms in the field, though differently evaluated by necessity. 

Although we developed the concept of our W-Net independently, we later found that \citet{xia2017wnet} had developed a concatenated U-Net and first used the name W-Net. \citet{GARE2022102326} also used the name W-Net for a multi-input U-Net, concatenating feature extraction at the network bottleneck. \citet{liu2023wnet} again used the name W-Net for a dual-path network involving a U-Net and a separate fully convolutional network bearing no resemblance to ours. So, while the name ``W-Net'' clearly does not have a standardized structure, we keep it for our implementation since it is appropriate descriptively. We also note that our W-Net has a substantially different innovation in its skip connections between the two U-Nets, whereas \citet{xia2017wnet} connected the two U-Nets at the output of the first to the input of the second. We actually developed a similar W-Net structure to theirs in the course of our development, stacking 2, 3, 4, and even 5 U-Nets in sequence. Without dwelling on the details of discarded models in our development process, we note that the multiple stacked U-Nets provided a very interesting sequence of background removal, essentially removing different general types of background (high-frequency, static, etc.) in stages. We found our W-Net to provide the more compact and effective solution in our case, but we suggest that the stacked U-Net may have future astrophysical application as a background subtraction method.

It is important to further discuss our development of the Adaptive Normalization method, which we consider to be a major innovation of our work. We noted in Section \ref{sec:scaling} that we found quantile transformation to be an effective method of scaling for TESS data, both in this and other work. It mutes the effect of the extreme flux values, generally caused by systematics or scattered light, and emphasizes the differences between fluxes in a range useful for predictive analysis, but it is not learnable. In most applications, this is not a concern. In the case of asteroid detection, however, the relevant flux value range is highly sensitive. Referring back to Figure \ref{fig:AN}, we note that the learned range of important values of the median-subtracted flux is approximately between -5 and 5 $e^{-}s^{-1}$. All other values approach zero or unity, with no discriminative capability as input to the W-Net, which will learn a rule something like ``input values near zero or one have no utility and should be ignored.'' In this manner, the Adaptive Normalization both emphasizes the flux region of interest and masks what is essentially noise.

We chose the Logistic CDF as the basis function for the Adaptive Normalization because of its simplicity. Two parameters define the distribution: The `location' ($\mu$) and the `scale' ($s$), to which we apply a learned weight ($w$). The learned values of these parameters (see Table \ref{tab:AN}) are instructive. Each $\mu$ value is strongly positive, near 1 $e^{-}s^{-1}$. Intuitively, this makes sense. An asteroid passing through a pixel increases the flux, usually imperceptibly from a visual perspective, hence why stacking (as in shift-and-stack) sequential images centered along an asteroid's path is valuable in revealing a slight flux increase in the central pixel. The Adaptive Normalization learns this and emphasizes it for follow-on processing by the W-Net.

In Section \ref{sec:can}, we noted that the Adaptive Normalization tends to learn one or two dominant CDFs that are strongly weighted, while others are weakly weighted. We acknowledge that ten CDFs were probably more than required to properly learn the scaling. However, we wanted to ensure that the adaptive scaling had the ability to learn its own distribution as needed, so we allowed for substantial flexibility. We take no stance on whether, in other potential applications of our Adaptive Normalization method, greater or fewer CDFs may be useful, and we suggest that this is highly dependent on the data and the problem to which it is being applied.

The primary difficulty in our work was the proper labeling of training data. While the labeling itself was cumbersome (Section \ref{sec:data_prep}), the downstream effects of labeling in model development (Sections \ref{sec:nn} and \ref{sec:loss}),  training (Section \ref{sec:training}) and even in evaluation (Section \ref{sec:results}), were decidedly nontrivial. We suspect that reliance on inherently incorrect training data would detract from any pixel-level ML-based analysis of asteroids.

With regard to positive pixel labels, they are both inherently uncertain (there exists uncertainty in the ephemeris which compounds with time since the last observation used in the ephemeris calculation as well as uncertainty in the TESS WCS solution) and subject to unexpected change (e.g.~gravitational alteration of the track). Negative pixel labels are plagued by the consequence of uncertain positive pixels (i.e.~a known asteroid's track has changed and it exists in negative-labeled pixels) as well as fundamental incompleteness (we have not identified nearly all asteroids, so there will be asteroid tracks present where they are not known to be). As we discussed in Section \ref{sec:completeness}, the net effect of these incorrect labels is that we are ``counter-training'' the model for a nontrivial fraction of pixels, that is, training true negatives as positives and true positives as negatives.

Considering quantitative evaluation of our performance, the labels present yet another challenge. Figures \ref{fig:roc_all_data} and \ref{fig:pr_all_data} aptly tell this story. Succinctly, if we consider all labels to be true and evaluate the model predictions against them, then our model is without merit (the left panels of Figures \ref{fig:roc_all_data} and \ref{fig:pr_all_data}). If we assume that the labels are incorrect when the model predicts a negative-labeled pixel with a score $>$0.999 or a positive-labeled pixel with a score $<$0.001 (i.e. the model is extremely certain one way or the other), then the model's quantitative performance is quite good (the right panels of Figures \ref{fig:roc_all_data} and \ref{fig:pr_all_data}).

Fortunately, we can assume that a substantial majority of our labels are correct, otherwise the model would not work. A neural network will always seek the best solution for the entirety of the training data, essentially overpowering a minority of incorrectly-labeled samples. This is not without harm, however. The incorrect labels create irregularities in the solution space that make it more difficult to train and harder to converge. So, while we achieved a successful result in spite of this obstacle, the model's performance was almost certainly not as good as it could have been.

As such, we have to wonder about the likelihood of improved model performance in the event that labels were perfect, or at least substantially improved. In retrospect, we suspect that a semi-supervised training scheme, whereby we alter the labels of the training data as training progresses, may have improved our outcomes. While we do not intend to repeat this exercise, we will implement such a training methodology (and suggest the same for others) in any future work involving this algorithm with asteroid data.  Alternatively, using synthetic data to train and validate separately from the real TESS data would have allowed for straightforward training and quantitative evaluation.  We suspect, however, that such a model would have struggled when applied to real data due to the difficulties of retrieving dim asteroid signals through the inherent TESS noise.  Again, we leave this investigation for future work.

While we examined the outputs of our model at the pixel level in Section \ref{sec:results}, our analysis ends at this point. The next logical step after our pixel level detection is, of course, track extraction. Comparison of discovered tracks to existing tracks can allow us to determine if our method made any novel detections. Although out of scope for this paper, we did begin this process. We are currently working with MPC on prerequisites and submission procedures and we anticipate eventually being able to evaluate newly discovered tracks and provide lightcurves and analysis in future work.

\section{Summary \& Conclusions}
\label{sec:conclusions}

We have presented the results of applying a custom neural network to the problem of asteroid detection in Years 1 and 2 of TESS data. To the best of our knowledge, this is the first pixel-level ML algorithm developed for asteroid detection. We developed a method of learned Adaptive Normalization as the first layer of the neural network in order for the noisy data to be optimally prepared for processing by the full network. Our neural network, which we call a W-Net, is two stacked U-Nets with skip connections between the U-Nets. We found our method to be robust to the difficulties of detecting moving objects combined with the intrinsic difficulty of TESS data itself.

We developed the publicly available \texttt{tess-asteroid-ml} package, for which we provide code, documentation, and tutorials. This code is the foundation of our work as it provides for the creation of training data by projecting all known ephemerides from the JPL Horizons database onto TESS FFIs. Subsequently, the code can divide stacked FFIs complete with asteroid track masks into memory-manageable pieces suitable for training ML algorithms.

We trained and tested two separate instances of our model: one trained/tested entirely on northern TESS sectors (14-26) to be used for inference on southern TESS sectors (1-13) and vice versa for the other model. In this manner, we ensured that there was no cross-contamination. We made the model robust to uncommon asteroid trajectories by augmenting training samples with all 16 possible rotations of the 3D data in equal probability.

We evaluated the model on all Year 1 and 2 TESS data, creating prediction scores that we compared against the track masks created by \texttt{tess-asteroid-ml}. We found that our training labels, though mostly correct, have a nontrivial fraction of incorrectly labeled pixels which are unavoidable due to the inherent uncertainty of the tracks, among other factors. Assuming our model was correct when it was extremely certain of the label, our model performance was excellent, attaining 0.887 ROC AUC (0.319 PR AUC) for all asteroids with $V<22$, improving to 0.956 ROC AUC (0.477 PR AUC) for all asteroids with $V<21$, 0.987 ROC AUC (0.620 PR AUC) for all asteroids with $V<20$, and 0.994 ROC AUC (0.661 PR AUC) for all asteroids with $V<19$.

We measured the completeness of our results with respect to the JPL Horizons asteroid database. Our detection results, though difficult to quantify at the pixel level, are similar to the results of \citet{2021PASP..133a4503W} and \citet{2024AJ....167..113N}. The model's 50\% detection magnitude is in the range 19.6 - 21.1, depending on the chosen prediction score threshold.

Most importantly, we demonstrated that a ML model can be developed for asteroid detection that does not require speed or trajectory assumptions.  All the work of filtering static background and finding moving objects in three dimensions at different speeds and trajectories is accomplished implicitly through the convolutional filters of the deep W-Net. Unlike shift-and-stack methods, our model is unbiased towards inclination and velocity vectors (besides the inherent bias of the training data, which can be corrected synthetically), meaning that we are able to identify an unbiased sample of asteroids. This will enable the study of asteroid populations using TESS with reduced observation bias, allowing us to understand the fundamental properties of asteroids in our solar system. While we designed this method for TESS data, it is certainly adaptable to any similar survey data.

The TESS dataset provides a valuable input sample of large, stable images with high cadence, uniform sampling. In this work we focused on early TESS data, but our model could be utilized on the more recent, fast cadence FFIs to find objects moving faster. Using our model on this dataset is likely to yield more information on asteroids nearer to the Earth.  TESS observes large portions of the sky for a month at a time, making it an underutilized workhorse for observing asteroids and NEOs. Using our model, TESS's survey can support our understanding and identification of asteroids. Our model could be used with any TESS-like data to support future endeavors to identify and track NEOs.  While the model would require training data, a similar architecture could be use to support missions such as Roman and NEO Surveyor.

\section*{acknowledgments}

This paper includes data collected by the TESS mission, which are publicly available from the Mikulski Archive for Space Telescopes (MAST). Funding for the TESS mission is provided by NASA's Science Mission directorate.
Resources supporting this work were provided by the NASA High-End Computing (HEC) Program through the NASA Center for Climate Simulation (NCCS) at Goddard Space Flight Center.
Funding for this work for J.M.P., A.T. and C.H. is provided by NASA grant 80NSSC20M0192. 
The material is based upon work supported by NASA under award number 80GSFC24M0006.

\facilities{
TESS}

\software{
{\tt Astrocut} \citep{astrocut},
{\tt Astropy} \citep{astropy:2013, astropy:2018, astropy:2022},
{\tt IPython} \citep{ipython},
{\tt Keras} \citep{keras},
{\tt Lightkurve} \citep{lightkurve},
{\tt lkspacecraft} \citep{Hedges_2025},
{\tt Matplotlib} \citep{matplotlib},
{\tt NumPy} \citep{numpy}, 
{\tt Pandas} \citep{pandas},
{\tt SciPy} \citep{scipy},
{\tt Tensorflow} \citep{tensorflow},
{\tt tess-asteroid-ml} \citep{tess-asteroid-ml},
{\tt tess-asteroids} \citep{tuson_2025_16332750},
{\tt tess-ephem} \citep{tess-ephem},
{\tt tesscube} \citep{Hedges_tesscube},
{\tt tesswcs} \citep{tesswcs}
}

\bibliography{refs}{}
\bibliographystyle{aasjournalv7}

\end{document}